\documentclass[preprint]{aastex}
\usepackage{amsmath}
\usepackage{hyperref}

\usepackage[english]{babel}
\usepackage{graphicx}
\usepackage{latexsym}
\usepackage{amsmath,amssymb}
\usepackage{natbib}
\usepackage{dcolumn}
\usepackage{bm}
\usepackage{calrsfs}
\usepackage{ulem}
\usepackage[usenames]{color}
\usepackage{xspace}

\newcommand{\nn}{\mbox{} \nonumber \\ \mbox{} }
\newcommand{\ba}{\begin{eqnarray}}
\newcommand{\ea}{\end{eqnarray}}

\newcommand{\NS}{neutron star}

\newcommand\eg{\textit{e.g.}}
\newcommand\cf{\textit{cf.\ }}

\def\be{\begin{equation}}
\def\ee{\end{equation}}

\begin{document}

\title{Radiation- and pair-loaded   shocks}

\author{Maxim Lyutikov\\
Department of Physics and Astronomy, Purdue University, \\
 525 Northwestern Avenue,
West Lafayette, IN
47907-2036}

\begin{abstract}
We consider the structure of mildly relativistic   shocks  in dense media, taking into account the  radiation  and pair loading, and diffusive radiation energy  transfer within the flow. 
For increasing shock velocity  (increasing post-shock temperature), the  first important  effect is the efficient energy redistribution by radiation within the shock that leads to the appearance of an isothermal jump, whereby the flow reaches the finial state through a discontinuous isothermal transition. The isothermal jump, on scales much smaller that the photon diffusion length, consists of a weak shock and a quick relaxation to the isothermal conditions. Highly  radiation-dominated shocks do not form isothermal jump. Pair production can  mildly increase the overall shock compression ratio to $\approx 10$ (from $4$ for matter-dominated shocks and  $7$ of the radiation-dominated shocks). 
\end{abstract}


\maketitle

\section{Introduction}
In many astrophysical settings  shocks  propagate through dense environment with mildly relativistic velocities, so that the post-shock  radiation pressure can exceed the thermal pressure \citep{2010ApJ...725...63B}.
 In addition, astrophysical shocks may heat up the plasma to temperatures where post-shock pair thermal production may become important.  For example, the discovery of  GRB170817A associated with gravitational waves event GW170817 \citep{2017ApJ...848L..13A} is best explained due to the emission from the  jet propagating with mildly relativistic velocities through a dense wind generated by the accretion torus  \citep{2017ApJ...848L...6L,pbm17,gnph17,2017arXiv171005897B}.  In this paper we address the structure of pair and radiation loaded shock transitions.

Several non-standard phenomena are expected in these types of shocks. First, a large fraction of the energy flux can be converted into radiation  and pairs, thus modifying the thermodynamic properties of plasma by both contributing to effective mass density and to pressure.  Secondly, energy redistribution within the shock may have profound influence on the shock properties. Especially important is photon diffusion - even if the energy density in photons is below the thermal energy density, the high photon mean free path may modify the shock considerably  \citep{ZeldovichRaizer}.

One of the key features of the radiation-modified shocks is the appearance of an isothermal discontinuity \citep{1910RSPSA..84..247R,LLVI,ZeldovichRaizer}. As the incoming flow is decelerated and compressed by scattering of the incoming radiation, its temperature  and compression ration increase. For sufficiently strong shocks the final temperature is reached before the final compression ratio. As a  result, an isothermal jump forms. 

Appearance of the isothermal discontinuity in relativistic shocks have been previously considered by 
\cite{1984PhFl...27.1991C}, though they did not take into account  pair production. Here we take pair production fully into account. 
  
 We assume that plasma is sufficiently dense so that radiation is trapped within the flow. The first most important effect is the diffusion of radiation - even if energy density and mass loading due to radiation (and of pairs) are small comparable to matter energy density,  the redistribution of energy within the shocked flow is important. This leads to the formation of isothermal jump,  \cite{LLVI,ZeldovichRaizer} and \S \ref{first}.
 (To clarify the notation, the  isothermal jump is different from highly radiatively cooled isothermal shocks; in our case the flow is energy conserving.)
  On the other hand,  for very strong highly radiation-dominated shocks the  pressure created by photons and pairs may dominate over the ion pressure. This introduces further complications: the isothermal jump can disappear,  \cite{ZeldovichRaizer} and \S \ref{Highly}.

 \section{Protons, radiation and pairs }
 
 \subsection{Protons, radiation and pairs at relativistic temperatures}
 In what follows we normalize the temperature to the electron temperature
 \be
 \theta\equiv T /( m_e c^2)
 \ee 
 and the number $n$ density to the dimensionless $\tilde{n}$ 
 \be
 {n}\equiv \tilde{n}/\lambda_C^3
 \ee
where $\lambda_C= \hbar/(m_e c)$ is the Compton wavelength.
Dimensionless number density of $1$ corresponds to 
 \ba &&
 n = 1 /\lambda_C^3 = 1.7 \times  10^{31} {\rm cm}^{-3}
 \nn &&
 \rho = m_p n = 2.8 \times 10^{7} {\rm g\, cm}^{-3}
 \label{param}
 \ea
 Such densities are expected in the region near the accretion shock during core collapse  \citep{2001PhRvL..86.1935M} and in the tidal tails/wind generated during \NS\ mergers \citep[\eg][]{2010MNRAS.406.2650M}.
 For example, in the case of \NS\ merger, a fraction of up to  $10^{-2}\, M_\odot$ is injected with mildly relativistic velocities  $v _w \sim 0.1 c$ for a duration  $\approx 1$ second. The corresponding dimensionless density  evaluates to  $ \tilde{n} \sim 10^{-5}$.  (Below we drop the tilde sign over the dimensionless density).

Consider next three contributions to the total enthalpy of plasma - from thermal motion of ions $w_p$, radiation $w_r$ and pairs $w_\pm$ (defined per unit volume). For  plasma ions and radiation,  the enthalpy can be written as
\ba  &&
\frac{\lambda _C^3}{c^2 m_e} w_p=\frac{\gamma  \theta  n}{\gamma -1}
\nn &&
\frac{\lambda _C^3}{c^2 m_e} w_r =\frac{4 \pi  \theta ^4}{45}
\ea

 The pair pressure, enthalpy and number density is a fairly complicated function of temperature, especially in a weakly relativistic regime \citep[\eg][and reference there in]{1979A&A....72..367W,1984MNRAS.209..175S}. To simplify the consideration, we adopt the following  parametrization
\ba  &&
\frac{\lambda _C^3}{c^2 m_e}w_\pm= \frac{\sqrt{2} {g_E} e^{-1/\theta } \sqrt{\theta ^3} (4 \theta +1)}{\pi
   ^{3/2}}
    \nn &&
 g_E =1+ 0.47 \theta +0.37 \sqrt{\theta }+\sqrt{\frac{2}{\pi }} \theta ^{3/2} \zeta (3),
 \label{gE}
    \ea
see     \cite{1979A&A....72..367W,1984MNRAS.209..175S}, $\zeta$ is the zeta-function
   (thus, pairs are assumed to be relativistic, factor of $4 \theta $ above). We expect that these simplifications for the pairs equation of state introduce only mild corrections.
   
   For plasma with ion number density $n$ (normalized to $\lambda_C^{-3}$) the ratios of ion mass density to that of pairs is  (for $\theta \leq 1$)
  \be
  \frac{m_p n}{m_e n_\pm} = \frac{\pi^{3/2}}{\sqrt{2}} \frac{ e^{1/\theta}}{ \theta^{3/2}} \mu n 
  \ee
    where $\mu = m_p/m_e$.
  So that pairs dominate ions by mass for 
  \be 
  n \leq  \frac{\sqrt{2} }{\pi^{3/2} }{ e^{-1/\theta}}{ \theta^{3/2}}  \mu^{-1} = 5 \times 10^{-5}
  \ee
   where the last equation assumes $\theta = 1$. (In physical units this corresponds to density of $ \rho \sim 1.4 \times 10^3$ g cm$^{-3}$.
   
    The ratios of ion mass density to that of radiation is
      \ba &&
  \frac{m_p n}{u_{rad}/c^2} = \frac{15}{\pi} \frac{\mu n }{ \theta^{4}}
  \nn &&
      u_{rad} = \frac{4}{c} \sigma_{SB} T^4 \equiv \frac{\pi}{15} \theta^{4}
  \ea
  Thus  the  radiation mass-density dominates   the ion density at much lower $n$,
       \be
  n \leq \frac{\pi}{15}\frac {\theta^3}{\mu}
  \ee
(corresponding  density is $ \rho \sim 3.2 \times 10^3$ g cm$^{-3}$ at $\theta=1$.
   
The  ratio of the  ion pressure to  radiation pressure is
 \be
 \frac{ nT}{u_{rad}/3} = \frac{45}{\pi} \frac{ n }{\theta^3}
\ee
 so that for 
   \be
  n \leq \frac{\pi}{15} {\theta^3}
  \ee
the  radiation pressure   dominates   the ion pressure (for $\rho < 2 \times 10^6$  g cm$^{-3}$).
   
 The pressure contribution of the radiation $p_r =u_{rad}/3 $ becomes dominate over matter contribution at mush higher densities: 
    \be
  \frac{ n T}{p_r} = \frac{45}{\pi} \frac{n}{ \theta^{3}} ,
  \ee
  so that pairs dominate ions by mass for 
  \be 
  n \leq  \frac{\pi}{45}{ \theta^{3}}  
  \ee
(this evaluates to $\rho = 2 \times 10^6$  g cm$^{-3}$ for  $\theta=1$).
  
   \subsection{Photons and pairs at mildly relativistic shocks}
   
The relations discussed above can be adopted to   shock-heated plasma. In this case for matter-dominated regime
   \be
   \theta \approx 2 \frac{\gamma-1}{(1+\gamma)^2} \mu \beta_1^2= \frac{3}{16} \mu \beta_1^2
   \ee
   If the upstream plasma has density $n_1$, 
   the post-shock ion pressure become smaller than radiation pressure at 
   \be
   \frac{2 \mu n_1 \beta_1^2 }{\gamma+1}< \frac{\pi}{45} \theta^4 
   \rightarrow
   n< \frac{8\pi(\gamma-1)^4}{45 (\gamma+1)^7} {\beta_1^6 \mu^3}
   \label{nnn}
   \ee
   In such  highly radiation-dominated shock without  pairs (still, mass density is dominated by ions)
  \ba && 
  \eta = 1/7
  \nn && 
  \theta_{max} = \left( \frac{270}{7 \pi} \right)^{1/4}  n_1^{1/4} \mu^{1/4} \beta_1^{1/2}
  \label{nopairs}
  \ea
  Also,  in the
  high compressibility limit $\eta \rightarrow 0$
  \be
  \theta = \left( \frac{45}{ \pi} \right)^{1/4}    n_1^{1/4} \mu^{1/4} \beta_1^{1/2}
  \ee
which is very close to  (\ref{nopairs}).

   For relativistic temperatures the condition  (\ref{nnn})  is also approximately  the condition for pair pressure to dominate over ion kinetic pressure, while at 
   $\theta \leq 1$ the post-shock ion pressure become smaller than pair pressure for 
   \be
   n_1 \leq \frac{4 \beta _1^3 (\gamma -1)^{5/2} \mu ^{3/2} e^{-\frac{(\gamma +1)^2}{2 \beta _1^2
   (\gamma -1) \mu }}}{\pi ^{3/2} (\gamma +1)^4}
   \ee
   Due to  exponential  dependence this ratio varies over large values  at mild shock velocities.
   
\begin{figure}[h!]
 \vskip -.1 truein
 \centering
\includegraphics[width=0.99\textwidth]{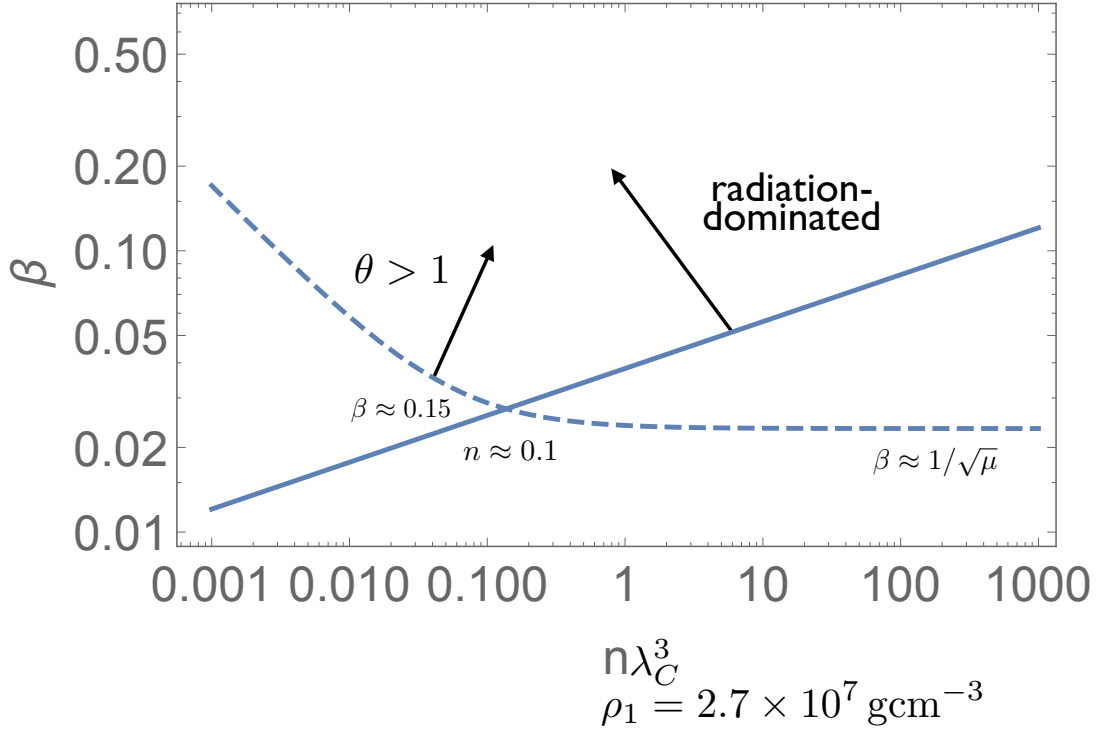}
\caption{Regime of radiation-dominated  shocks and relativistic post-shock temperatures  on the $\beta-n$ diagram, see Eqns. (\protect\ref{beta1}) and   (\protect\ref{beta2})}
\label{conditions}
\end{figure}

   Rewriting  Eq. (\ref{nnn}) as a condition on velocity, the post-shock radiation domination requires 
   \be
   \beta > \left( \frac{60}{\pi} \right)^{1/6} n ^{1/6} \mu^{-1/2}
   \label{beta1}
   \ee
   The post-shock temperature becomes $\theta > 1$ at  
   \be
   \beta \geq  \frac{1}{\sqrt{\mu}}\sqrt{1 + \frac{\pi}{60 n}},
    \label{beta2}
   \ee
  (this relation is valid for both matter and radiation-dominated shocks),  see Fig. \ref{conditions}
   
\section{Stationary 1D flow with radiation and pair production}
\label{mmain}

 Let's assume that a shock propagates  with velocity $\beta_1 \leq 1$ through a cold plasma with ion density $\rho_1$. 
  At each point one-dimensional, stationary non-relativistic flow is described by the following set of equations (mass, momentum and energy flux conservation)
   \ba &&
  \beta_1 \rho_1 = \beta  \rho
  \nn &&
   \rho_1 \beta_1^2 = p_{ tot} + \rho_{ tot} \beta^2
   \nn &&
    \rho_1 \beta_1^3/2=( w_{ tot} +  \rho_{ tot} \beta^2/2) \beta +F_r
    \label{conserv}
    \ea
        where subscript $1$ indicates  quantities far upstream.
        
In (\ref{conserv})
    \be
      \rho_{tot} = \rho+ n_\pm m_e +u_{rad}/c^2
      \ee
       is the total mass density that includes ion density $\rho$, pair mass density $n_\pm m_e$ and effective mass density of radiation $u_{rad}/c^2$ ($u_{rad}$ is the energy-density of thermal radiation),
  \be
   w_{2} = \frac{4}{3}   u_{rad} + \frac{\gamma}{\gamma-1} \frac{\rho}{m_p} T +  n_\pm (m_e c^2 +  4 T)
   \ee
   is the plasma enthalpy not including ion rest mass,     that includes contribution from radiation, ion energy density, pair rest mass and pair energy density, $\gamma =5/3$ is the adiabatic index of the ion component (assumed non-relativistic), the electron component is assumed to be relativistic (factor of $4$ in the expression for  the pair contribution). 
   Finally, $F_r$ is the energy flux   carried by radiation. We assume high optical depth limit, so that locally
\ba &&
  F_r =  - \frac{c}{3 n_{ tot}\sigma_T} \nabla      u_{rad} 
     \nn && 
      u_{rad} = \frac{4}{c} \sigma_{SB} T^4
      \label{1}
\ea
where total density has contributions both from neutralizing electrons and pairs, $n_{ tot} = {\rho}/{m_p} + n_\pm$.

\section{The first effect: formation of isothermal jump in strong shocks}
\label{first}

As the velocity of the shock grows, the  energy redistribution within the flow by radiation becomes the first important effect - even if the radiation pressure is much smaller than the matter pressure.
Let's neglect radiation energy density and pressure, but keep energy diffusion due to photons with large mean free path \citep[\cf][]{ZeldovichRaizer}. Everywhere along the stationary flow we have conservation of matter, momentum and energy flux:
\ba &&
  \beta_1 \rho_1 = \beta  \rho
  \nn &&
   \rho_1 \beta_1^2 = \frac{\rho}{m_p} T + \rho \beta^2
   \nn &&
    \rho_1 \beta_1^3/2=( \frac{\gamma}{\gamma-1} \frac{\rho}{m_p} T+  \rho \beta^2/2) \beta +F_r
       \label{111}
     \ea

Introducing the inverse of the compression ratio
$
\eta = n_1/n
  $, 
  Eq. (\ref{111}) becomes
  \ba &&
\theta= \mu \beta _1^2 (1-\eta ) \eta  
\label{31}
\\ &&
\frac{\lambda _C}{15 \beta _1 n_1^2 \alpha _f^2}\partial_x \theta= \frac{\beta _1^2 \left(\eta ^2-1\right) \mu +\frac{2 \gamma  \theta }{\gamma -1}}{\eta 
   \theta ^3}
   \label{33}
\ea
Far downstream temperature becomes constant, $\partial_x \theta=0$, and (\ref{31}-\ref{33}) give shock jump conditions
\ba &&
\eta_2= \frac{\gamma-1}{\gamma+1}
\nn &&
\theta_2=\frac{2 \beta _1^2 (\gamma -1) \mu }{(\gamma +1)^2}
\ea

Importantly,  in order to reach this final state a flow should develop a special type of discontinuity - isothermal jump \citep{LLVI,ZeldovichRaizer}.
The isothermal jump in this case forms regardless of a particular form of  $F_r$, as we discus next.

The momentum conservation  (\ref{31}) can also be written for  the evolution of compression as a function of temperature $\eta(\theta)$
\ba &&
\eta= \frac{1}{2} \left(1\pm  \sqrt{1-\frac{4 \theta }{\beta _1^2 \mu }} \right) = 
\frac{1}{2} \left(1\pm  \sqrt{1-\frac{\theta }{\theta _{{max}}}} \right)
\nn &&
\theta _{{max}}=
\frac{\beta _1^2 \mu }{4}
\label{21}
\ea
Thus, there are two branches for $\eta(\theta)$; they  connect at the  point  
\ba &&
\eta_{crit}=\frac{1}{2} 
\nn &&
\theta _{max}= \frac{\beta _1^2 \mu }{4}
\ea
Importantly, the initial point $\eta=1$ is on the upper branch, while  the final jump conditions is on the lower branch. Yet the flow cannot pass continuously from the initial to the final point - 
using (\ref{31}) the equation for the compression ratio becomes
\be
\frac{\pi  \beta _1^5 (\gamma -1) \mu ^3}{15 {\alpha _f}^2 {n_1}^2 \lambda _C^5}
\partial_x \eta =
\frac{\gamma  (\eta -1)+\eta +1}{(1-2 \eta ) (1-\eta )^2 \eta ^4}
   \label{deta1}
   \ee
      Equation (\ref{deta1}) clearly shows that the final state of $\eta =\eta_2 =  1/4 $ (for $\gamma=5/3$) cannot be reached  continuously: in order to reach it continuously  one must pass through a special point $\eta =1/2$. This statement is true for any non-zero left hand side.
      
\begin{figure}[h!]
 \vskip -.1 truein
 \centering
\includegraphics[width=0.99\textwidth]{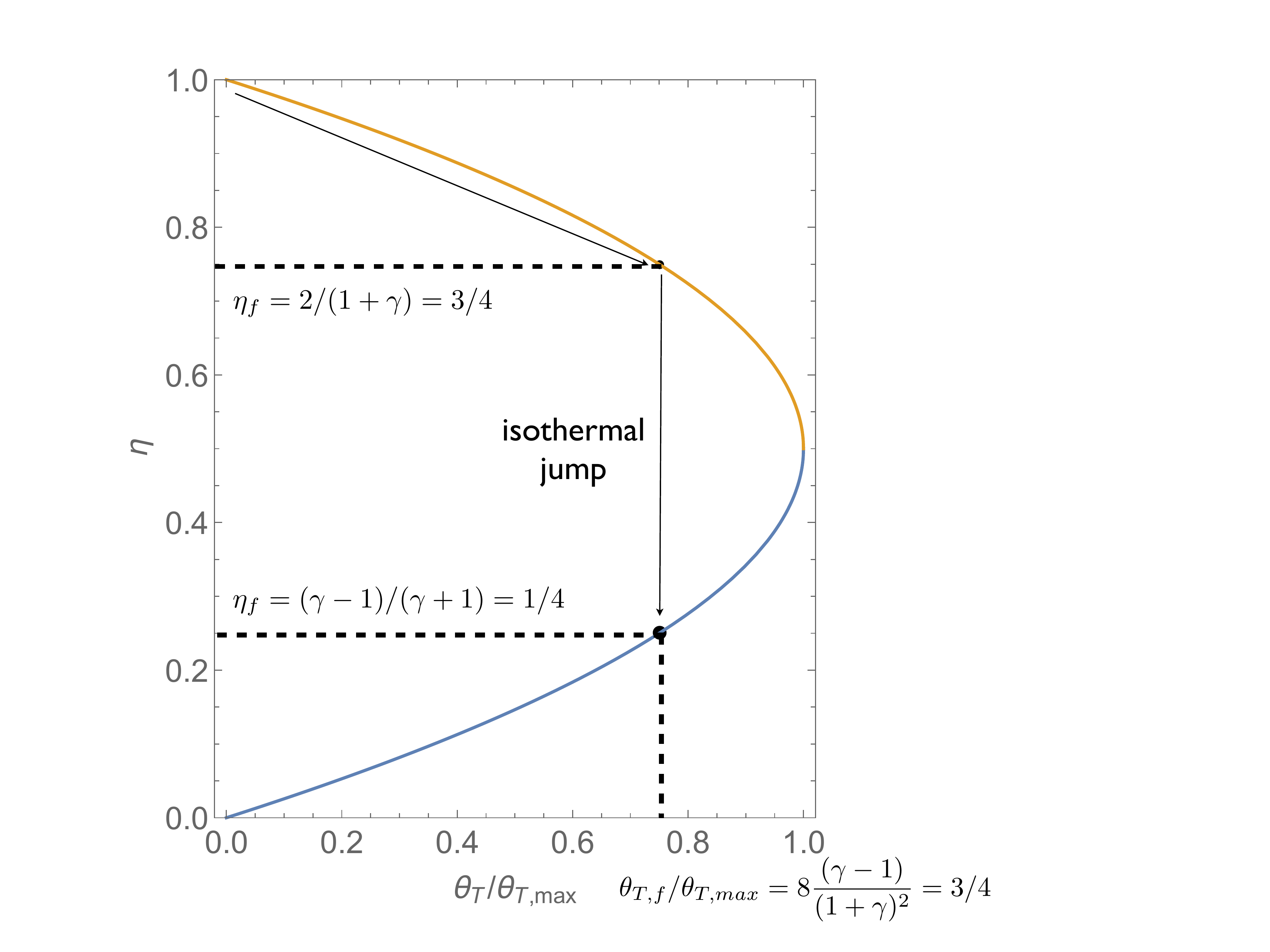}
\caption{Evolution of the initially cold flow within the radiatively-mediated shock  in the  compression ratio  $\eta$  --  temperature $\theta$ plane neglecting momentum flux of pairs and radiation. The flow starts at $\eta=1, \, \theta=0$ and  reaches the finite temperature $\theta_2$ at the moment when the inverse compression ratio $\eta_+$  is smaller than the final $\eta_f$. At this point the flow experiences an isothermal jump to the final state.  Two highlighted  points correspond to the isothermal  jump condition. 
On scales smaller than the photon mean free path a weak fluid sub-shock results supersonic-subsonic transition than then smoothly evolves toward the final state.
}
\label{ofbeta00}
\end{figure}

          Note, that
\be
\frac{\theta_2}{\theta_{max}} = 8 \frac{\gamma-1}{(\gamma+1)^2}= \frac{3}{4} < 1
\ee
Thus, as the state evolves along the upper branch, the terminal temperature is reached before the the terminal compression. It is required that temperature increase monotonically \citep[\eg][ Eq. (95.3)]{LLVI}. Thus, since  ${\theta_2}<{\theta_{max}}$, the final state cannot be reached continuously. 
There should be  an isothermal jump at $\theta=\theta_2$, Fig. (\ref{ofbeta00}).
\footnote{The local Mach number  is 
\be
M =  \sqrt{\frac{ \eta}{\gamma(1-\eta)} },
\ee
it equals unity at $\eta_{M=1} = \gamma/(1+\gamma)  =5/8$ (at this point $\theta = \gamma/(\gamma+1) \mu \beta _1^2 =(15/64) \mu \beta _1^2 $).
}

 \subsection{Structure of the precursor}
 
\begin{figure}[h!]
 \vskip -.1 truein
 \centering
\includegraphics[width=0.99\textwidth]{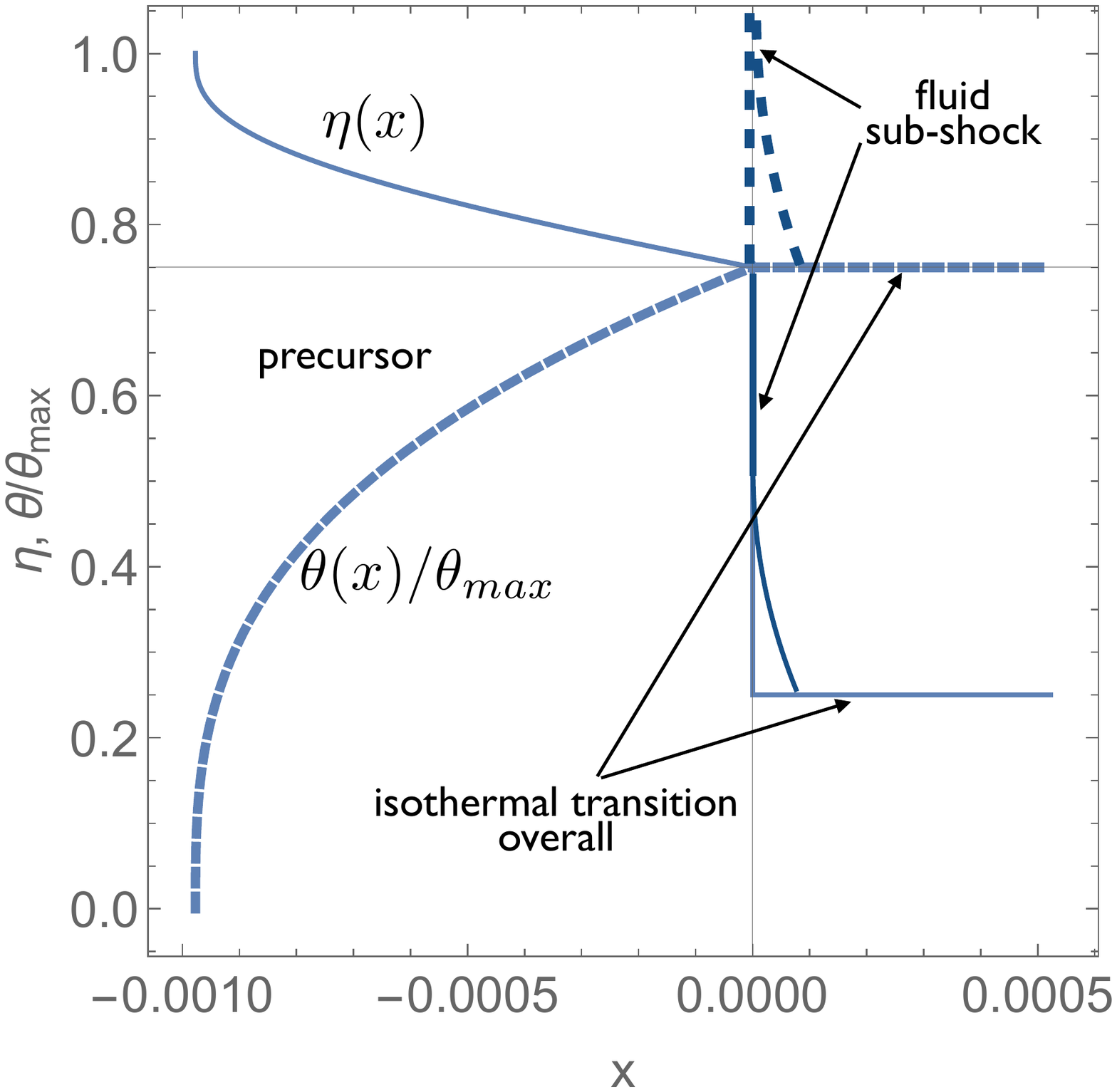}
\caption{ Compression ratio  and temperature evolution within an isothermal transition, analytical solution of Eq. (\ref{prec}).  The precursor extends to a  finite distance in front of the isothermal transition (for cold upstream); distance is normalized to $L_p$, Eq. (\protect\ref{Lp}). In the precursor the  flow is heated and decelerated by the radiation pressure   Near the front of the precursor located at $x=x_p\approx -0.00097$ the compression ratio is  $\eta =1 -(6 (x-x-p))^{1/3}$. The isothermal jump is   located at $x=0$, where the final temperature $\theta_2= (3/4) \theta_{max}$  and compression $\eta =1/4$ are reached.  On scales much smaller than the radiative diffusion length the isothermal jump consist of weak sub-shock and a narrow relation layer. For $\gamma = 5/3$ the subshock heats the plasma to the maximal possible temperature of $\theta_{max}= \mu \beta_1^2/4$.}
\label{ofbeta}
\end{figure}

  In fact we can find an exact analytical solution of (\ref{deta1}). 
 Let us normalize out, by  the precursor scale
 \be
 L_p = \frac{15 {\alpha _f}^2 {n_1}^2 \lambda _C^5}{\pi  \beta _1^5 (\gamma -1) \mu ^3}
 \ee
The  precursor scale $L_p$ is  larger than the mean free path $\lambda _{mfp} = 1/(\sigma _T n_1)$,
 \be
 \frac{ L_p }{\lambda _{mfp} } = \frac{8}{45} \pi ^2 \beta _1^5 (\gamma -1) \mu ^3
 \label{Lp}
 \ee
 for $\beta \geq 10^{-2}$.
 We find
\be
\partial_x \eta =
\frac{\gamma  (\eta -1)+\eta +1}{(1-2 \eta ) (1-\eta )^2 \eta ^4}
\label{prec}
\ee
which has an analytical solution for $x(\eta)$ with boundary condition $x(1)=0$, 
   Fig. \ref{ofbeta} (temperature is related to compression ratio via (\ref{21}) along the upper branch). 
   Eq. (\ref{prec}) describes the precursor of the isothermal transition under assumption that energy density and pressure  of pairs and radiation can be neglected, yet radiative flux diffusively redistribues energy within a flow.

 Near $x\approx 0$ the compression ratio and temperature is  $ \eta \propto x^{1/3}$ (so that $d\eta/dx$ diverges). This is  a drawback of the  neglect of the upstream temperature; finite shock Mach number leads to the appearance of an exponential precursor, Appendix \ref{precursor}.

\subsection{Appearance of the iso-thermal jump}

The above derivation assumed that the upstream medium is cold, so that the shock is infinitely strong. If the upstream plasma has temperature $\theta_{1}$ (so that Mach number is $M_1 = \sqrt{ \frac{\mu}{\gamma \theta_{1}}} \beta_1$), the compression ratio is 
\be
\eta_\pm =\frac{1}{2} \left(\frac{\theta_{1}}{\beta_1^2 \mu}-\sqrt{\left(\frac{\theta_{1}}{\beta_1^2
   \mu}+1\right){}^2-\frac{4 T}{\beta_1^2 \mu}}+1\right)
  \label{etapm}
\ee
Thus, the maximal temperature is 
\be
\theta_{max}= \frac{\left(\beta_1^2 \mu+\theta_{1}\right){}^2}{4 \beta_1^2 \mu}=\frac{\beta_1^2 \mu \left(\gamma  M_1^2+1\right){}^2}{4 \gamma ^2 M_1^4}
\ee
Post-shock temperature  and compression ratios are
\ba &&
\theta_2 = -\frac{2 (\gamma -1) \gamma  \theta_{1}^2}{(\gamma +1)^2 \beta_1^2 \mu}+\frac{2 (\gamma -1) \beta_1^2
   \mu}{(\gamma +1)^2}-\frac{\left(\gamma ^2-6 \gamma +1\right) \theta_{1}}{(\gamma +1)^2}
   \nn &&
   \eta_2=\frac{(\gamma -1) \beta_1^2 \mu+2 \gamma  \theta_{1}}{(\gamma +1) \beta_1^2 \mu}
   \ea
Equating $\theta_{max}$ to $\theta_2$ we find that isothermal jump forms for 
\be
\theta_{1} < \frac{(3-\gamma ) \beta_1^2 \mu}{3 \gamma -1}, \, 
M_1 > M_{crit} =  \sqrt{\frac{3 \gamma -1}{(3-\gamma ) \gamma }}= \frac{3}{\sqrt{5}}= 1.34
\label{Mcrit}
\ee
At this point $p_2/p_1= (\gamma+1)/(3-\gamma)$, \cf, \cite{LLVI} Eq. (95.7).

Note that the ratio of the final temperature $\theta_2$ to maximal temperature $\theta_{max}$ never exceeds unity:
\ba &&
\frac{\theta_2}{\theta_{max}}= 
\frac{-8 (\gamma -3)^2 (\gamma -1) \gamma +4 \left(3 \gamma ^4-28 \gamma ^3+66 \gamma ^2-28
   \gamma +3\right) m_1^2+8 (1-3 \gamma )^2 (\gamma -1) m_1^4}{(\gamma +1)^2 \left(-\gamma
   +(3 \gamma -1) m_1^2+3\right){}^2} \leq 1
   \nn &&
   m_1= \frac{M_1}{M_{crit}}
   \ea
   This ratio reaches unity only at  $m_1=1$. In this case 
   \be
   \eta _{crit}= \frac{1+ \gamma}{3 \gamma-1} = 2/3
   \ee
Overall properties of the isothermal jump for strong shocks are further discussed in Appendix  \ref{properties}.

\subsection{Internal structure of the isothermal jump}
The appearance of the isothermal jump is a mathematically oddity, related to the diffusive approximation  and the assumption of the local thermodynamics equilibrium. This prevent the formation of the temperature jump. 
Also,  
On the scales smaller than the photon mean free path the radiation becomes decoupled from the plasma. As a result a fluid subshock forms with a typical thickness much smaller than the photon mean free path.

 Let us consider the properties of this fluid subshock. It will occur when the temperature of the flow reaches the final temperature $\theta_2$ , while the local compression ratio is $\eta_+$. The corresponding Mach number 
 (upstream of the fluid subshock)  is
 \be
 M_{s,1}= \frac{\eta_+}{\sqrt{\gamma \theta_2}} = \sqrt{\frac{2}{\gamma(\gamma-1)}}  = \frac{3}{\sqrt{5}} =1.34
 \label{Ms}
 \ee
 (for $\gamma =5/3$, and only for this $\gamma$, $M_s$ equals $M_{crit}$, Eq. \ref{Mcrit}). The sub-shock is weak. 

A subshock with Mach number (\ref{Ms}) has  (the inverse) compression ratio $\eta_{s}$
\be
 \eta_{s} = \gamma-1 = 2/3
 \ee
 The corresponding temperature jump is 
 \be
 \frac{\theta_{s}}{\theta_2} =( 3 -\gamma)  =  \frac{4}{3} 
 \ee
while  the post-shock temperature is
 \be
 \theta_{s,2} =\frac{2 (3-\gamma ) (\gamma -1)}{(\gamma +1)^2} \mu \beta^2_1 =\frac{1}{4} \mu \beta^2_1
 \ee
   Thus, for $\gamma=5/3$ the subshock reaches the maximal possible temperature. 
   
 The total compression ratio from the upstream is 
 \be
  \eta_{s,2} = \eta_+ \eta_{sub} = 2  \frac{\gamma-1}{\gamma+1} =1/2,
  \ee
see Fig. \ref{ofbeta00}

The post-fluid shock Mach number is 
\be
M_{s,2} = \sqrt{\frac{2 (\gamma -1)}{(3-\gamma ) \gamma }} = \sqrt{3/5}
\ee
 
 \section{Highly radiation-dominated shocks: no isothermal jump}
 \label{Highly}
 Above in \S \ref{first} we considered the simplest most that demonstrates the formation of the isothermal jump. We neglected the contribution of radiation and pairs to mass and energy density (and pressure), but allowed for the diffusive  energy redistribution within the flow by long range photon propagation.

  Next, let us consider another extreme case of  strongly radiation-dominated shocks, when pressure and energy density of matter can be neglected in comparison with radiation, yet when inertia is dominated by matter.
Thus, we are in the regime
\be 
\frac{\pi \theta^4}{15 \mu}  \ll n \ll \frac{\pi \theta^3}{45} 
\ee

 The flow is then described by the  system of conservation laws
    \ba &&
  \beta_1 \rho_1 = \beta  \rho
  \nn &&
   \rho_1 \beta_1^2 = p_{rad} + \rho \beta^2
   \nn &&
    \rho_1 \beta_1^3/2=( w_{rad} +  \rho \beta^2/2) \beta +F_r
 \nn &&
   w_{rad} = \frac{4}{3}   u_{rad} =4 p_{rad}
   \nn &&
   F_r = - \frac{4}{3} \frac{ a c}{(\rho/m_p) \sigma _T} T^3 \partial_x T
   \label{noloading}
   \ea
We find
 \ba &&
 \eta = 1 - \frac{6}{7} \left(\frac{\theta}{ \theta_{ max}} \right)^4
 \nn &&
\frac{ \mu  \lambda_ C \beta_1}{\alpha_f^2 } \partial_x \theta=
\frac{7 \pi}{135} \, \left(  \frac{ 1-  (\theta/ \theta_{ max})^4)}{1- (6/7) (\theta/ \theta_{ max})^4} \right)  \theta\theta_{ max}^4
\nn &&
\theta_{ max} = \left( \frac{270}{7\pi} \right)^{1/4}\mu^{1/4} n_1^{1/4}\beta_1^{1/2}
\label{radnopair0}
\ea
The $\eta-\theta$ dependence in this case is strikingly different from the case where pressure is dominated by matter (but transfer by radiation), \S \ref{first}. Now, compressibility evolves smoothly with temperature
until a final state
\ba &&
\theta_2= \theta_{ max}
\nn &&
\eta _2 =1/7
\ea
 is reached. 
The typical scale of the shock is $\Delta x \sim \mu  \lambda_ C \beta_1/\alpha_f^2 $ (transition layer becomes thicker with higher velocity.).

\begin{figure}[h!]
 \vskip -.1 truein
 \centering
\includegraphics[width=0.49\textwidth]{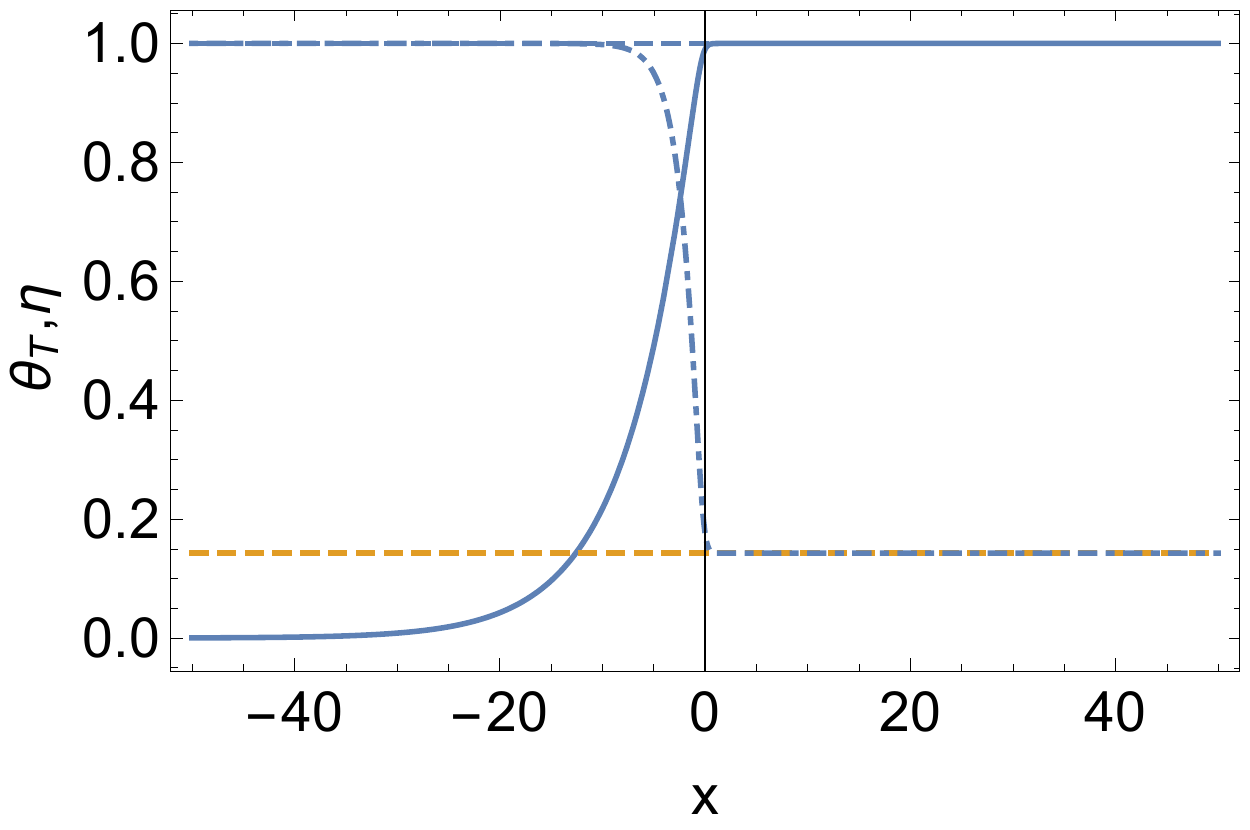}
\caption{Structure of the radiation-dominated shock with no pair formation, scattering-dominated. For this plot $\theta_{ max}=1$. Solid line is $\theta$, dot-dashed line is $\eta$. Coordinate is normalized to $ \mu  \lambda_ C \beta_1/\alpha_f^2$. The flow smoothly reaches the finite state without a discontinuous transitions.
}
\label{ofbeta0}
\end{figure}

Equation (\ref{radnopair0}) has an analytical solution (after dimensionalizing)
\be
x = \frac{135}{196 \pi} \theta_{ max}^{-4} \ln \left( \frac{\theta^{28}}{\theta_{ max}^4-\theta_{T}^4} \right) 
\ee
Fig. \ref{ofbeta0}.
 
Thus, for highly radiation-dominated shocks there is no isothermal jump. Mildly radiation-dominated shocks do have isothermal jumps as we demonstrate next, \S \ref{mild}

\subsection{Mildly radiation dominated shocks without pairs: reappearance of isothermal jump}
\label{mild}

Let us keep radiation energy density and matter contribution to pressure. In this case illuminating analytical results can be obtained in the limit $\beta_1^2 \rightarrow 0$
(but we keep $\beta_1^2 \mu $ terms.) The momentum and energy conservation give
\ba &&
\pi  \eta  \theta ^4+45 n_1 \left(\beta _1^2 (\eta -1) \eta  \mu +\theta \right)=0
\nn &&
\frac{\lambda _C}{\beta _1 n_1 \alpha _f^2} \partial_x \theta
=\frac{8 \pi  (\gamma -1) \eta  \theta ^4+45 n_1 \left(\beta _1^2 (\gamma -1) \left(\eta
   ^2-1\right) \mu +2 \gamma  \theta \right)}{3 (\gamma -1) \eta  \theta ^3}
   \label{nopairr}
   \ea
   
\begin{figure}[h!]
 \vskip -.1 truein
 \centering
\includegraphics[width=0.49\textwidth]{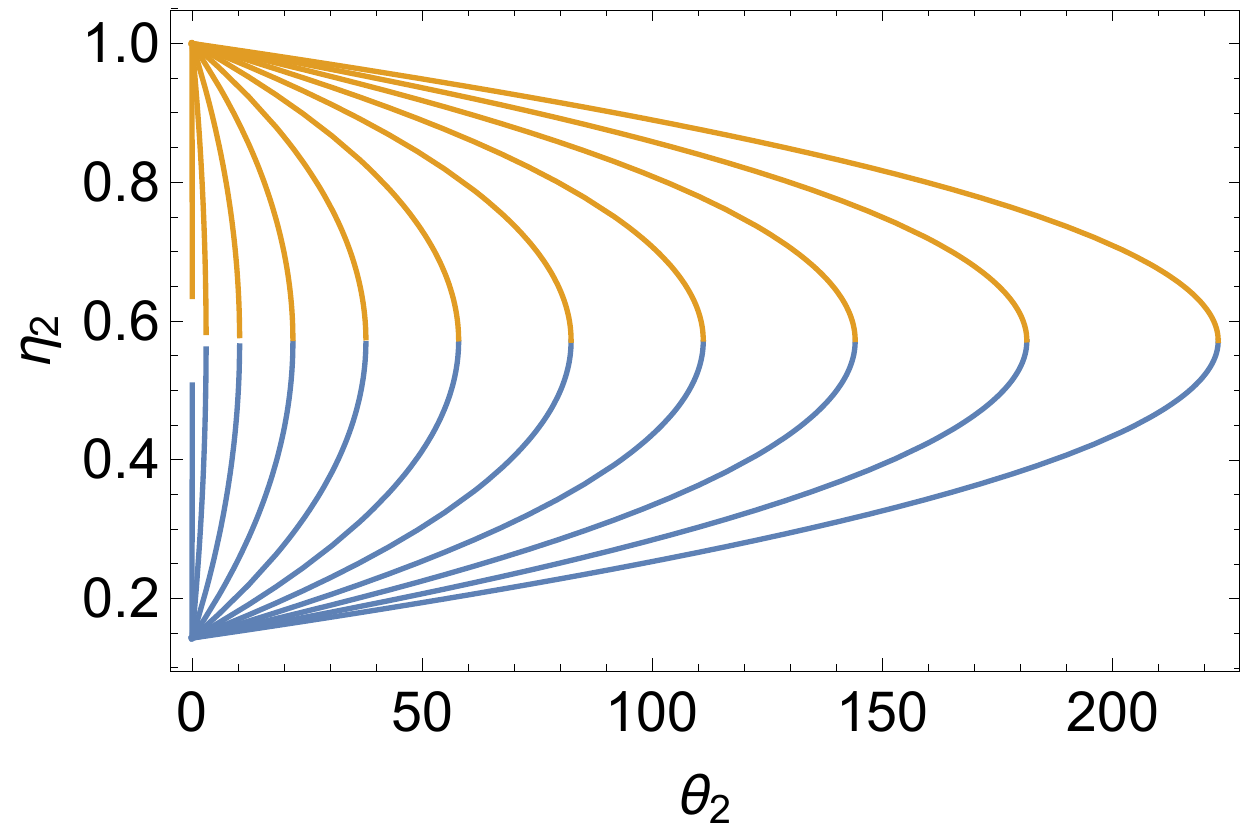}
\caption{Dependence of $\eta _2(\theta_2)$ for different $\beta_1 = 0.01, 0.06...0.51$ in case of no pairs. Each curve corresponds to a given $\beta_1$; different points on the curve correspond to different $n_1$. Only lower parts of the curves correspond to shock transitions. For each $\theta_2 < \theta_{max}$ the final state is reached via vertical isothermal jump between points on the curves.
}
\label{EtaofTbetaNopair1}
\end{figure}

   Condition $ \partial_x \theta=0$ corresponds to far downstream. In this case we can eliminate $n_1$ and find the relation between the final compression ratio and temperature (for a given $\beta_1$), Fig. \ref{EtaofTbetaNopair1}
   \be
   \eta_2 =\frac{1}{7}  \left(4 \pm {\sqrt{9   +(56-42 \gamma ) \frac{\theta_2}{\mu \beta _1^2 (\gamma -1)} }}\right)
   \ee
    For $\theta_2 \rightarrow 0$ this gives $\eta_2 = 1/7$ - compression ratio for strongly radiation-dominated shock (another root is the trivial $\eta_2=1$).

\begin{figure}[h!]
 \vskip -.1 truein
 \centering
\includegraphics[width=0.49\textwidth]{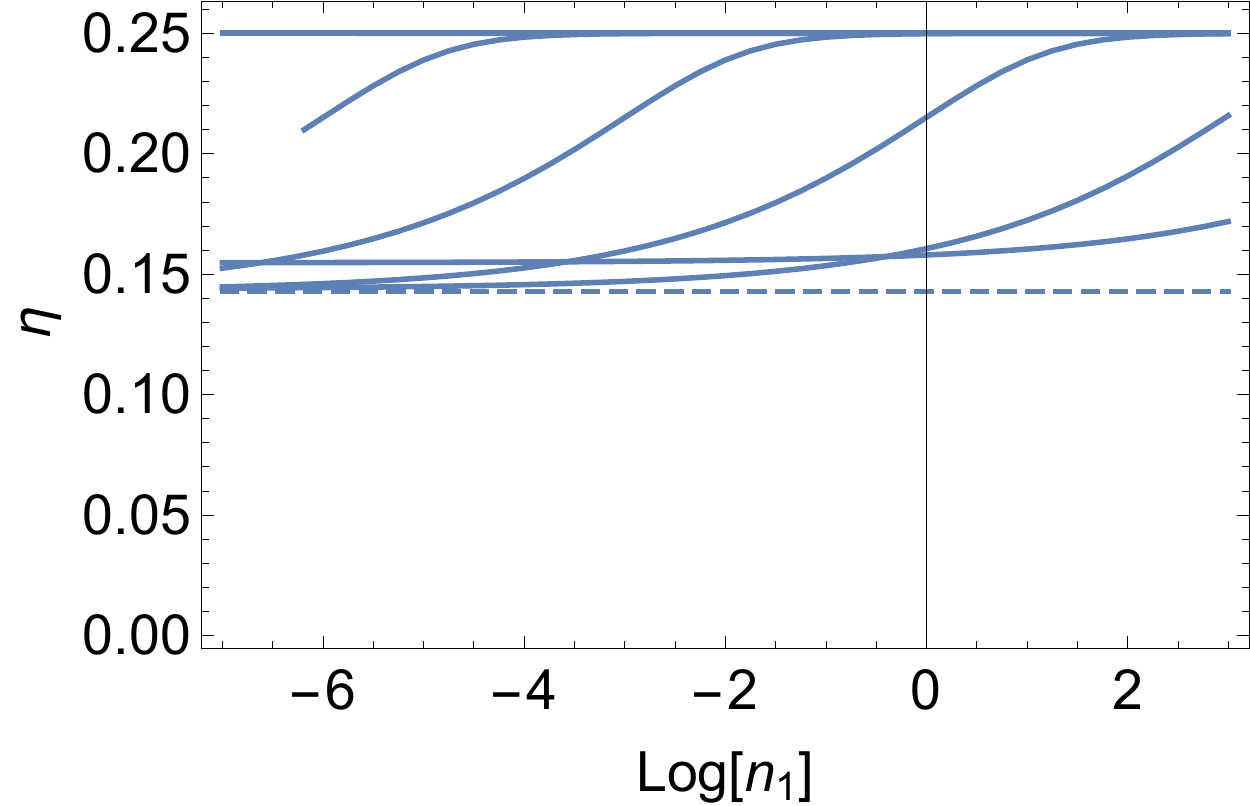}
\includegraphics[width=0.49\textwidth]{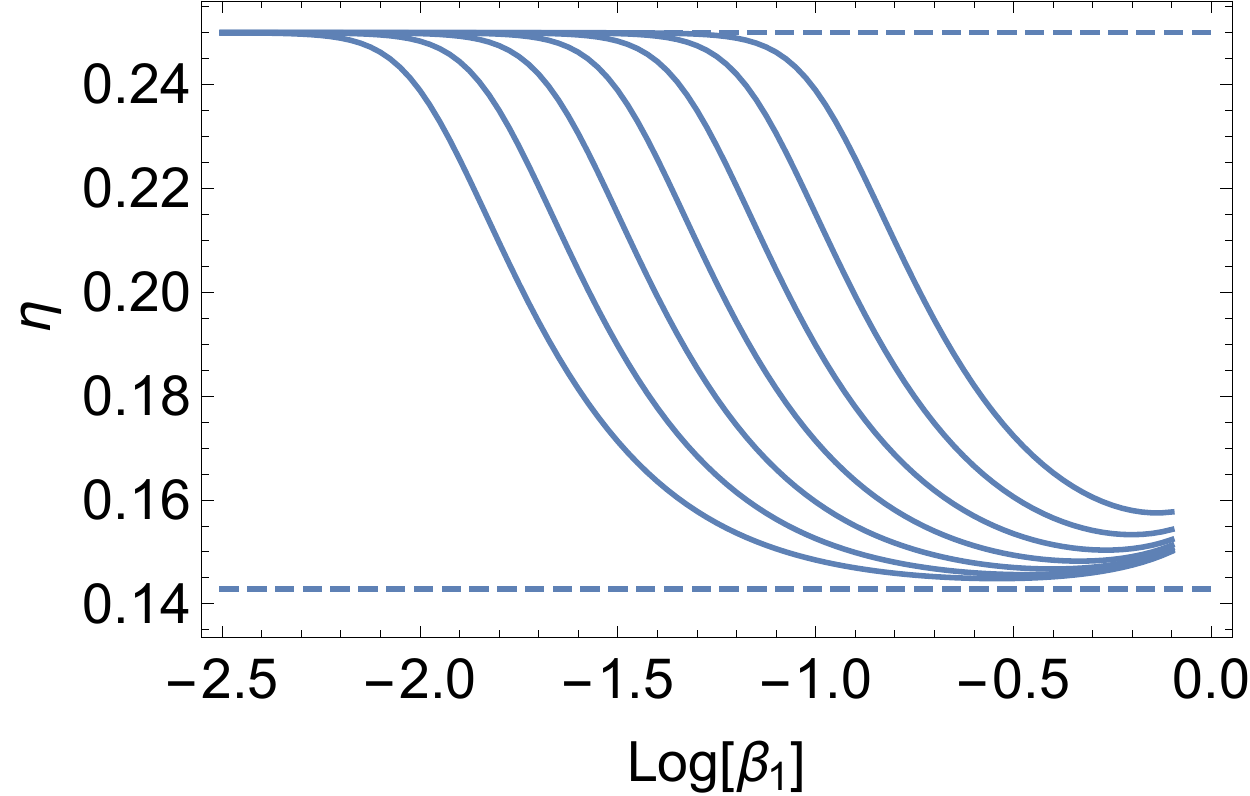}
\caption{Compression ratio as function of shock  velocity and upstream density; no pair creation. For high upstream density or low velocity the shock is matter-dominated with compression ratio approaching $1/4$, while for small density/high velocity the shock is radiation-dominated with  compression ratio approaching $1/7$. The compression ratio never goes above $7$.
}
\label{ofbetaNopair}
\end{figure}

We can also solve for compression ratio as a function of upstream parameters, Fig. \ref{ofbetaNopair}. Compression ratio never goes below $1/7$. 

\begin{figure}[h!]
 \vskip -.1 truein
 \centering
\includegraphics[width=0.49\textwidth]{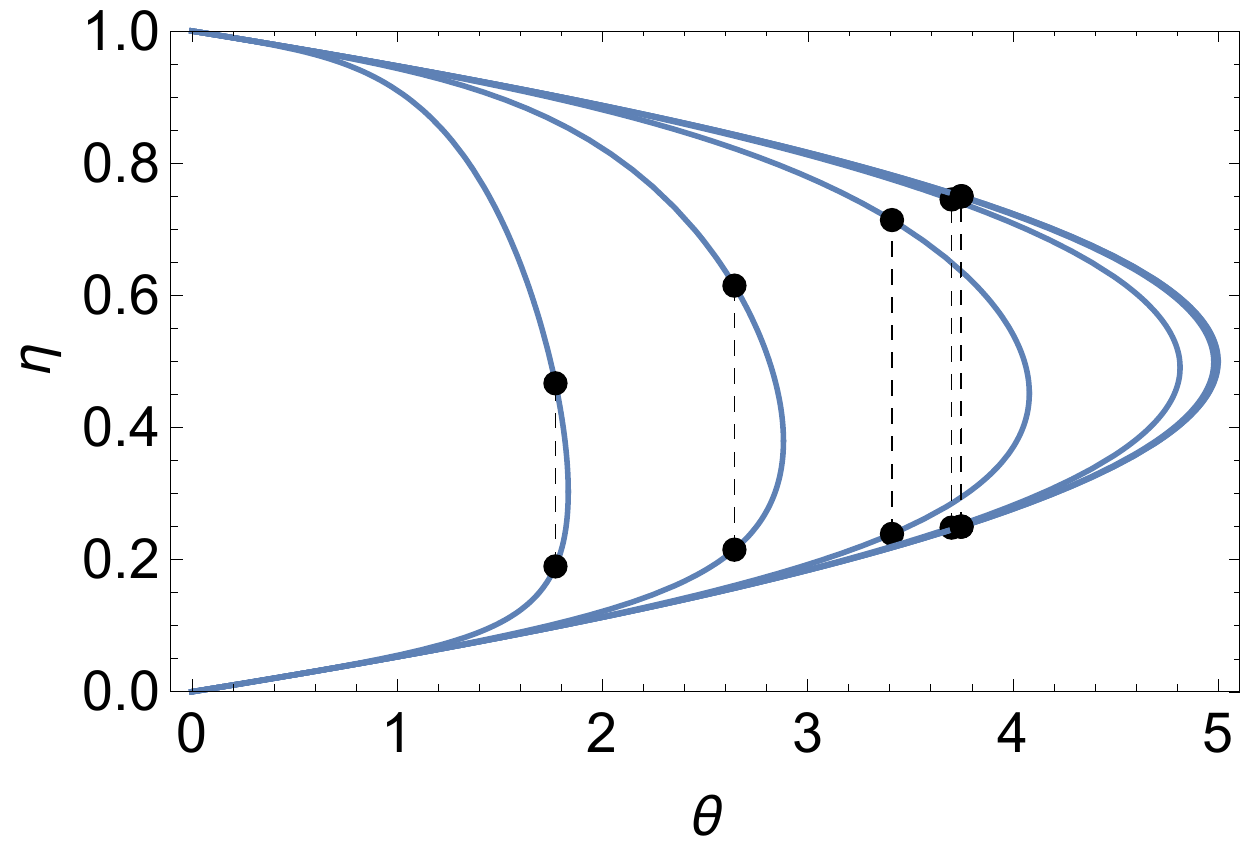}
\includegraphics[width=0.49\textwidth]{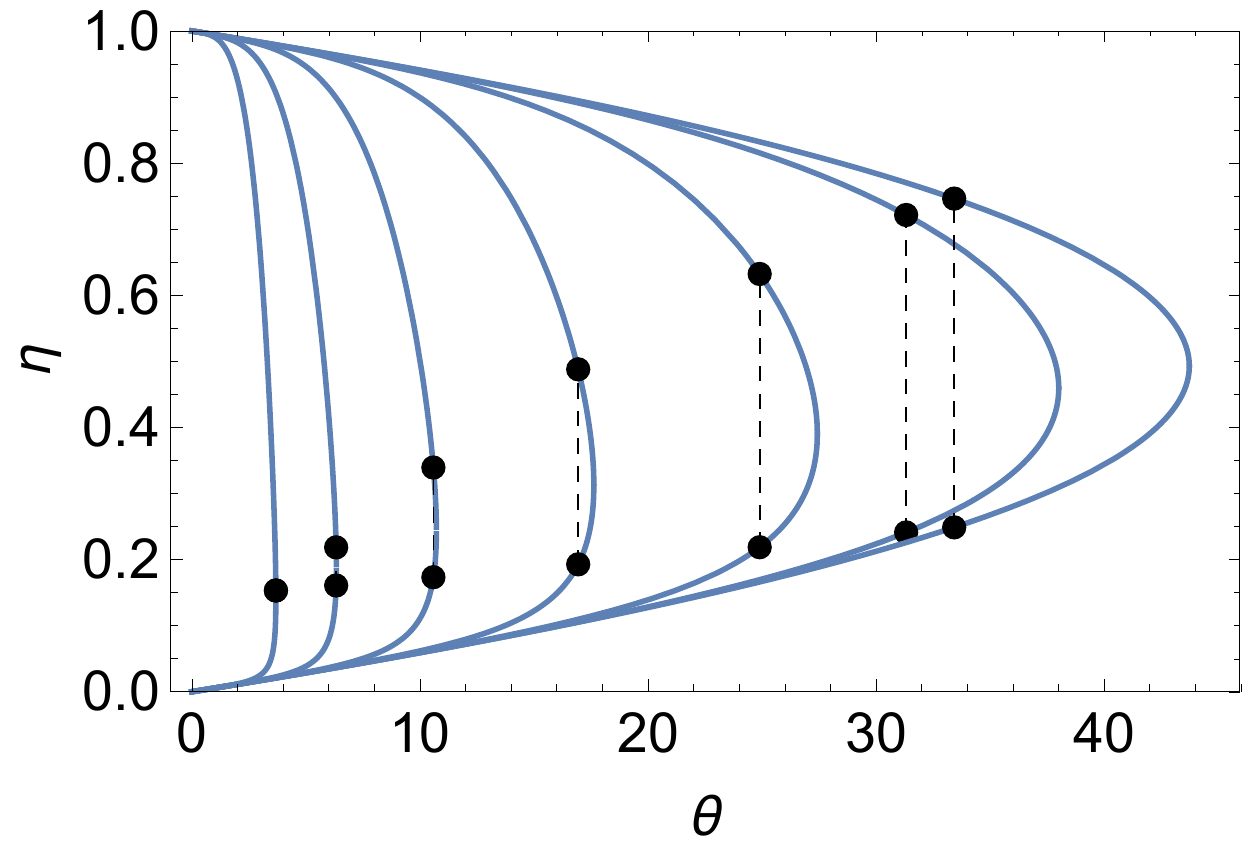}
\caption{Evolution of flow without pair production on $\eta-\theta$ diagram. The flow evolves  from the initial point $\eta=1, \, \theta=0$, the final temperature is reached at smaller compression ratios, upper dots. The isothermal jump forms, dashed lines. Different curves correspond to different densities, $n= 10^{-1}, \, 1...\, 10^5$. Left panel: $\beta_1 =0.1$, right panel  $\beta_1 =0.3$. For sufficiently small densities the isothermal jump disappears.
}
\label{EtaofTbetaNopair}
\end{figure}

Finally, let us investigate how the final state can be reached. Here the situation is very similar to the simplest case considered in \S \ref{first}. As the flow evolves on the $\eta-\theta$ diagram from the initial point $\eta=1, \, \theta=0$, the final temperature is reached at smaller compression ratios. As a result, isothermal shock forms, Fig. \ref{EtaofTbetaNopair}.
According to results in \S \ref{Highly}, for sufficiently small densities/high velocities the isothermal jump should disappear. This is indeed seen in Fig. \ref{EtaofTbetaNopair}, right panel.

  \subsection{The general case: radiation mediated shocks with pair production}
 After getting experience with simple cases we are ready to tackle the full problem:  flow with radiation and pair production
 We again make non-relativistic approximation, neglecting $\beta_1 ^2 \ll1$, but not when it comes with $\mu \gg 1$.  The momentum and energy conservation equations give
\ba &&
45
   \pi ^{3/2} n_1 \left( \theta - \beta _1^2 (1-\eta ) \eta  \mu \right) + 45 \sqrt{2} \eta  \sqrt{g_E} e^{-1/\theta } \theta ^{5/2}+\pi ^{5/2} \eta  \theta ^4=0
   \label{moment}
   \\ &&
  \frac{3 \pi ^{3/2} \eta  \theta ^3 \lambda _C}{\beta _1 \alpha _f^2}  \partial_x \theta = \left(\sqrt{2} \eta  g_e e^{-1/\theta } \theta ^{3/2}+\pi ^{3/2} n_1\right) 
  \times
  \nn &&
   \left(\frac{90 \sqrt{2} \eta  g_e e^{-1/\theta } (4 \theta +1) \theta ^{3/2}}{\pi
   ^{3/2}}+8 \pi  \eta  \theta ^4+\frac{45 n_1 \left(\beta _1^2 (\gamma -1) \left(\eta
   ^2-1\right) \mu +2 \gamma  \theta \right)}{\gamma -1}\right)  
    \label{main}
      \ea
   In the momentum equation terms are plasma contribution, pairs and  radiation consequently.

\begin{figure}[h!]
 \vskip -.1 truein
 \centering
\includegraphics[width=0.49\textwidth]{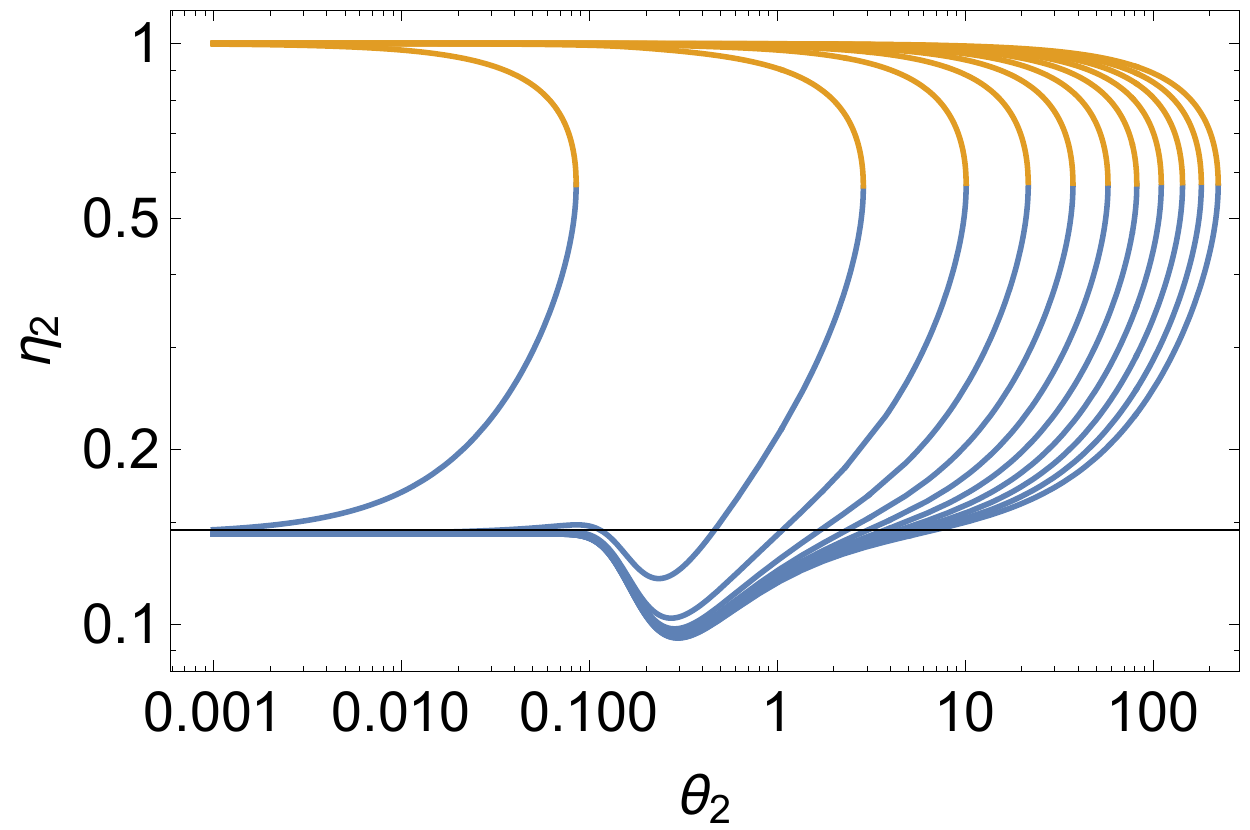}
\caption{ Dependence of $\eta _2(\theta_2)$ for different $\beta_1 = 0.01, 0.06...0.51$. Each curve corresponds to a given $\beta_1$; different points on the curve correspond to different $n_1$. Only lower parts of the curves correspond to shock transitions. For each $\theta_2 < \theta_{max}$ the final state is reached via vertical isothermal jump between points on the curves. }
\label{eta-theta}
\end{figure}

Setting $\partial_x \theta =0$ in (\ref{main}) determines the overall shock jump conditions.  Since (\ref{main}) is linear in $n_1$,  we can eliminate $n_1$ and find how the compression ratio far downstream depends on the final temperature (for fixed $\beta_1$), Fig. \ref{eta-theta}

Few points are worth mentioning. At mild post-shock conditions the compression ratio can go below $1/7$  and reach $\sim 1/10$. This fairly mild modification is due to the formation of pairs - energy is spent on creating mass, not pressure. The effect is fairly  mild since at smaller temperature there are few pairs, while at larger temperature pairs behave like radiation, so that their mass is not important. There is also a limiting case for large $n_1$, where pairs and radiation are 
not important, see (\ref{21}).

\begin{figure}[h!]
 \vskip -.1 truein
 \centering
\includegraphics[width=0.49\textwidth]{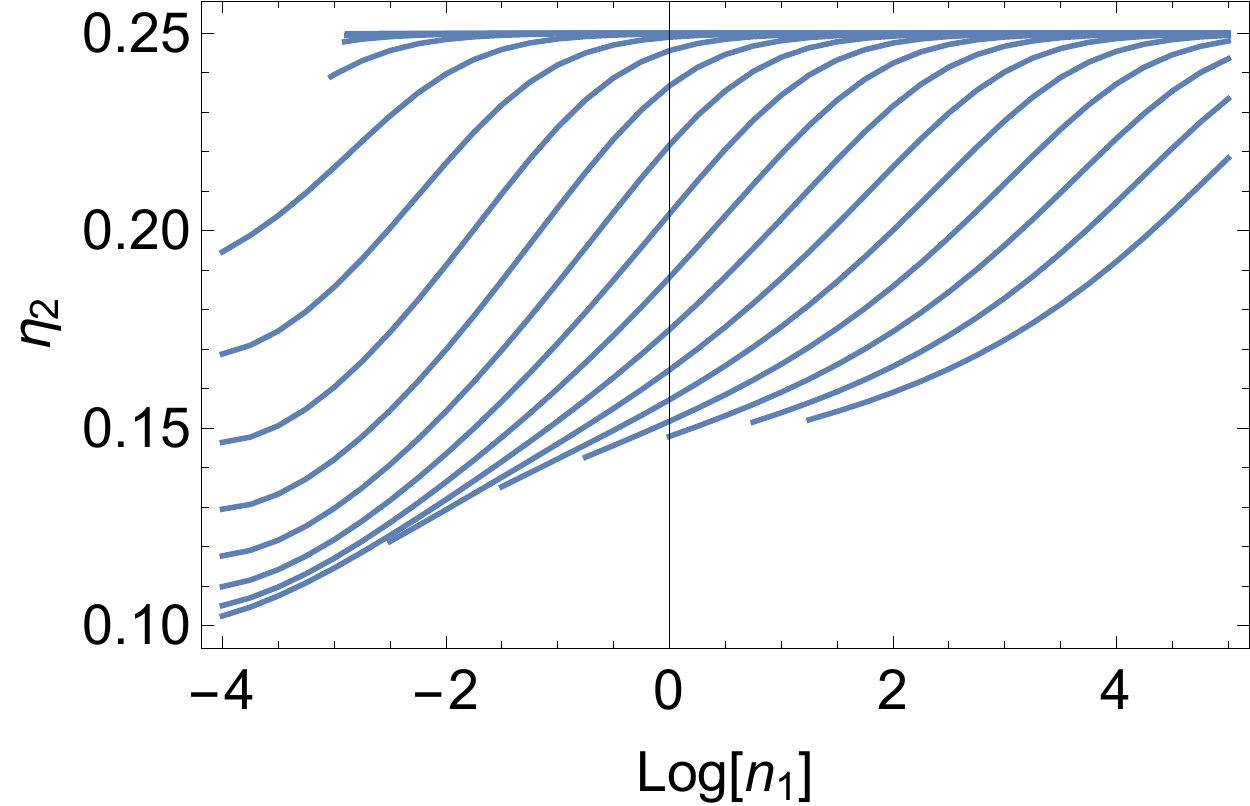}
\includegraphics[width=0.49\textwidth]{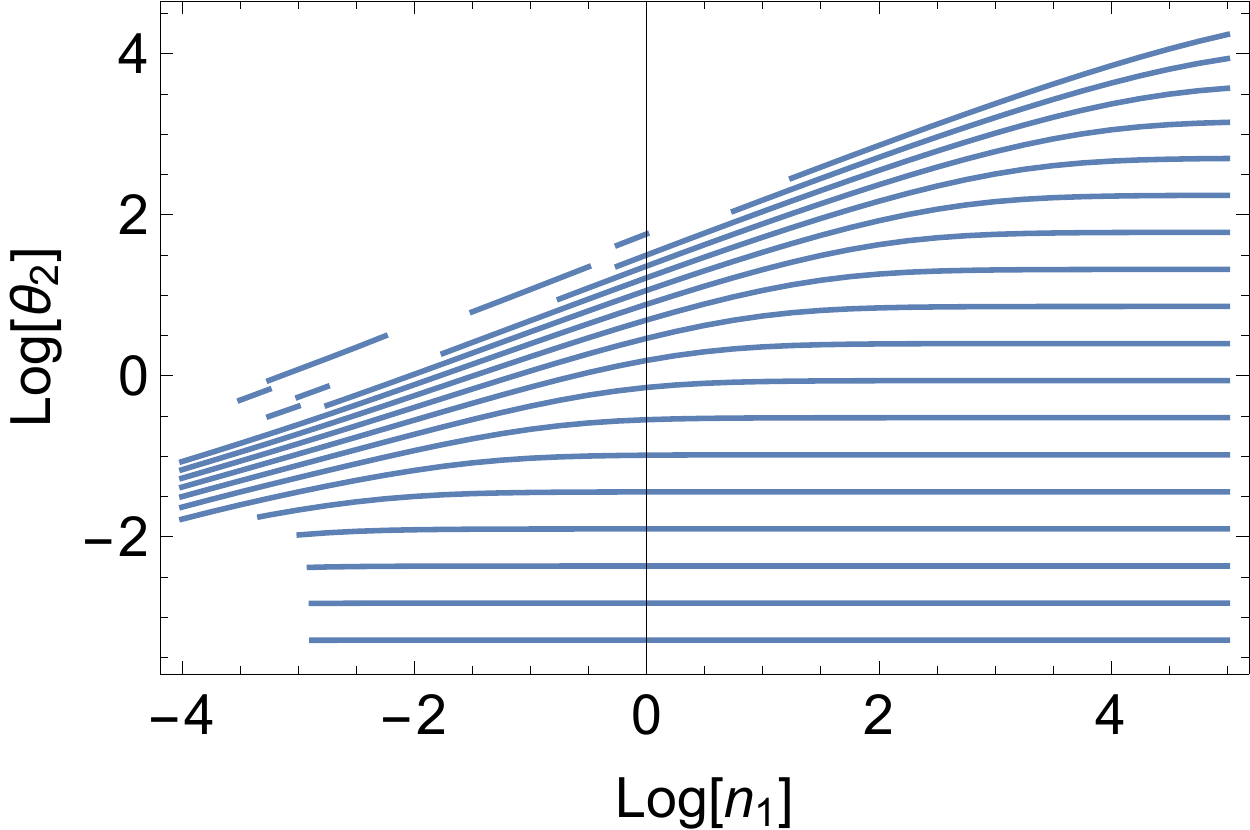}\\
\includegraphics[width=0.49\textwidth]{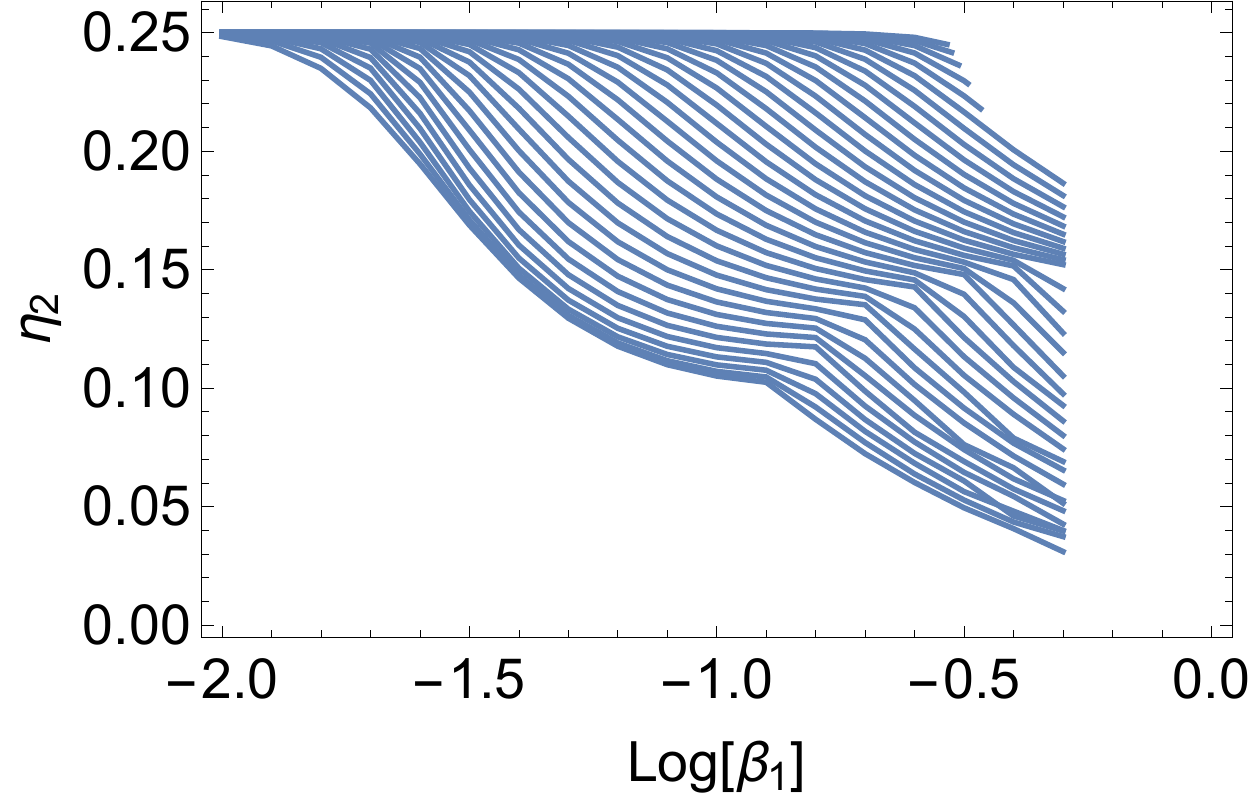}
\includegraphics[width=0.49\textwidth]{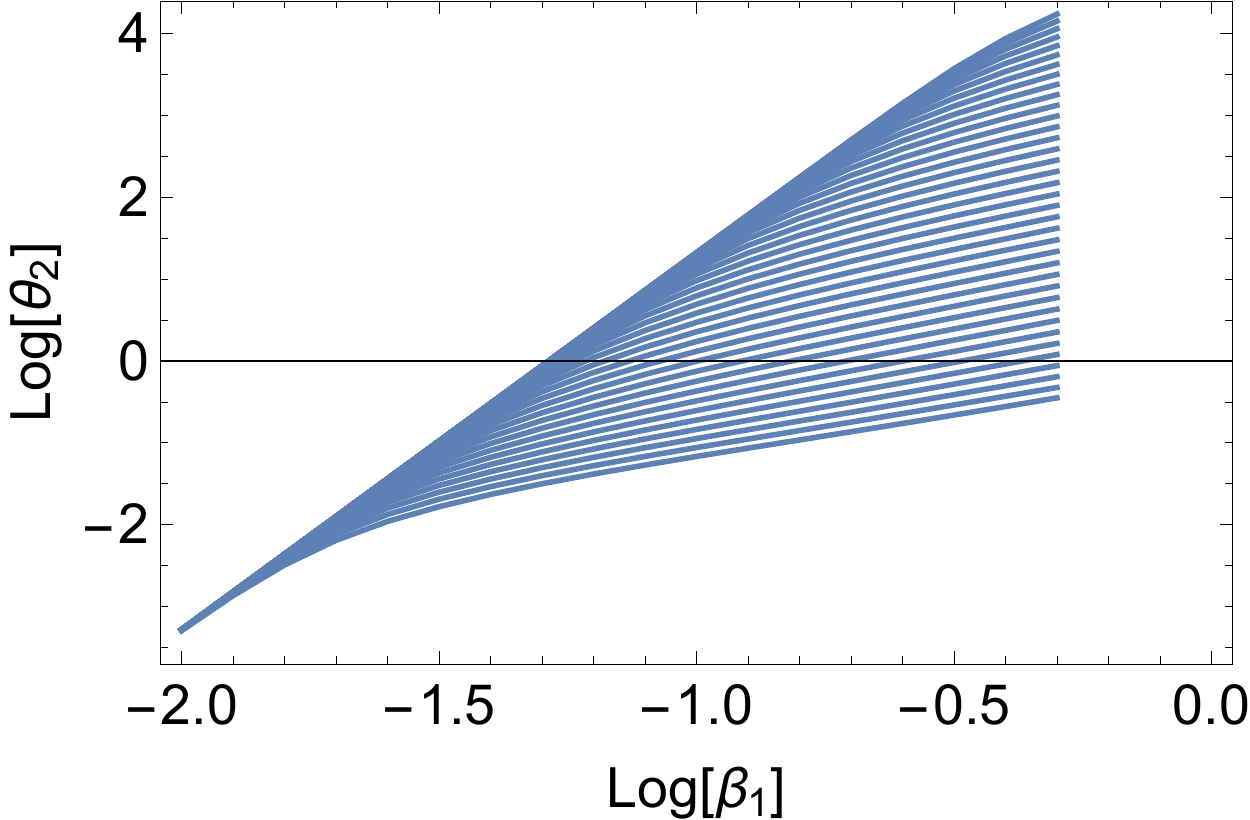}
\caption{Jump conditions  $\eta_2$ and $\theta_2$ for pair and radiative loaded shocks as function of upstream density for different shock velocities (top row, for $\beta _1 = 10^{-2},\, 10^{-1.75},...,  \, 10^{-0.25}$  ) and   as function of
shock velocity  for different upstream density (bottom row, $n_1 = 10^{-4}, \,  10^{-3.75}, ...10^{4}$). }
\label{eta-theta1}
\end{figure}

The shock jump conditions are plotted in Fig. \ref{eta-theta1}, where  the final  compression ration $\eta_2$ and the final temperature $\theta_2$ are plotted as functions of the upstream density and shock velocity

    \begin{figure}[h!]
 \vskip -.1 truein
 \centering
 \includegraphics[width=0.49\textwidth]{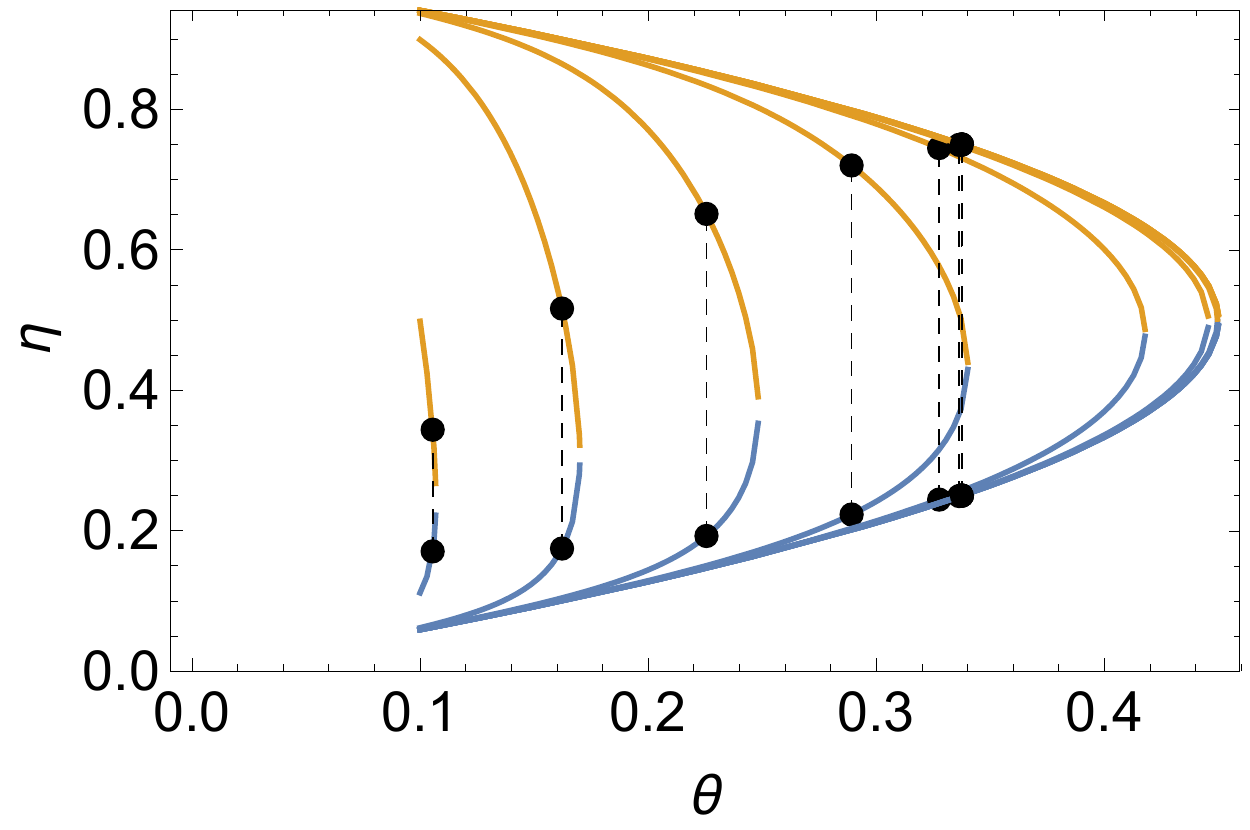}
\includegraphics[width=0.49\textwidth]{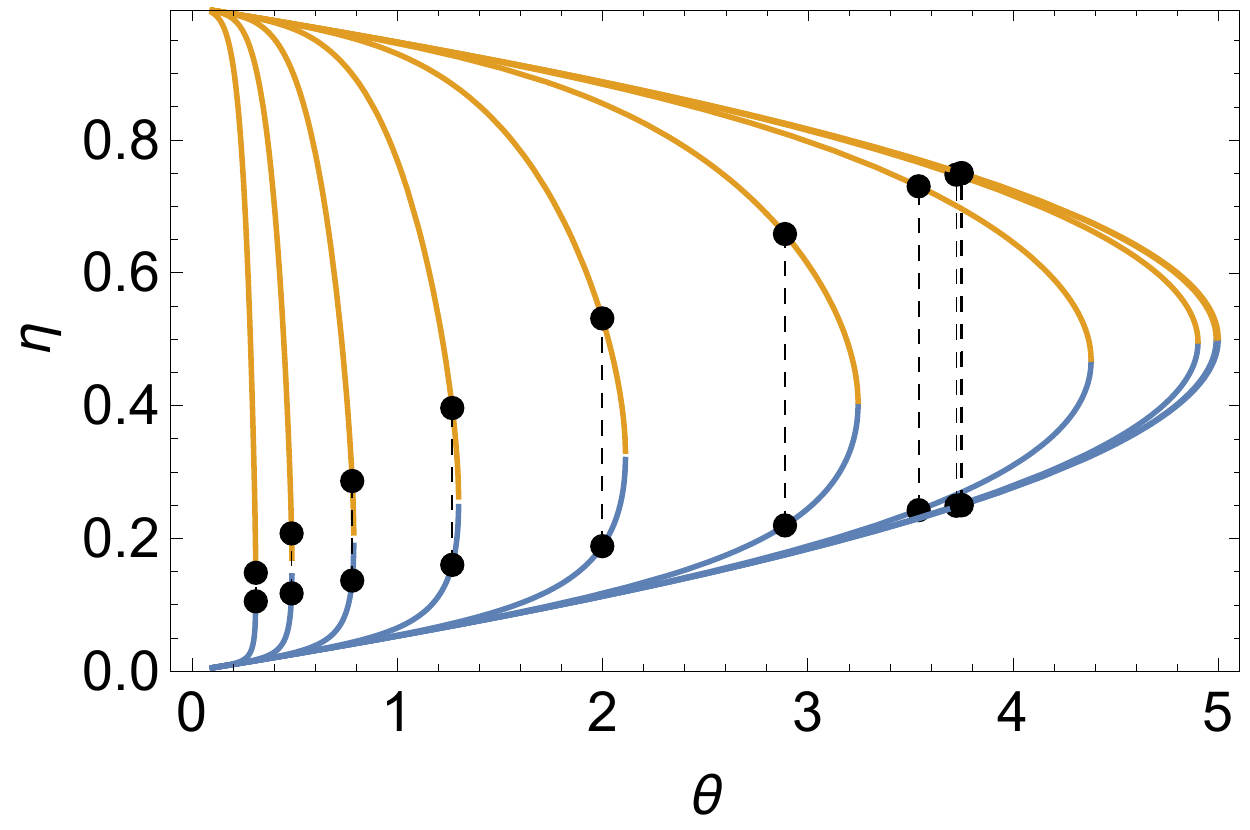}
\caption{Evolution of pair and radiative loaded flows on $\eta-\theta$ plane for $\beta_1=0.03,\, 0.1$, $n_1 =  10^{-4}, \, 10^{-3}... 10^{5}$.}
\label{eta-theta2}
\end{figure}

    \begin{figure}[h!]
 \vskip -.1 truein
 \centering
 \includegraphics[width=0.49\textwidth]{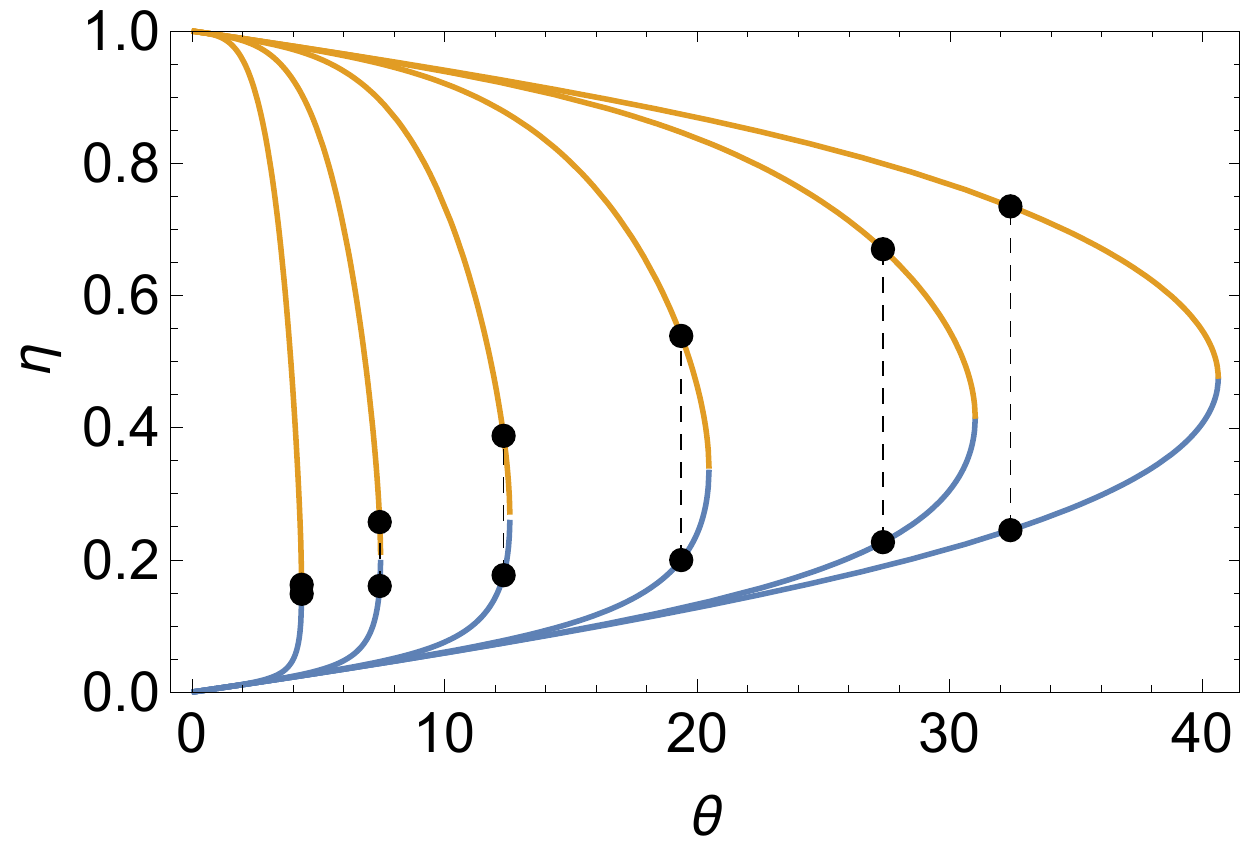}
\caption{Evolution of pair and radiative loaded flows on $\eta-\theta$ plane for $\beta_1=0.3$, $n_1 =  1,\,10... 10^{5}$. For small densities the  isothermal jump disappears.}
\label{eta-theta3}
\end{figure}

\begin{figure}[h!]
 \vskip -.1 truein
 \centering
 \includegraphics[width=0.49\textwidth]{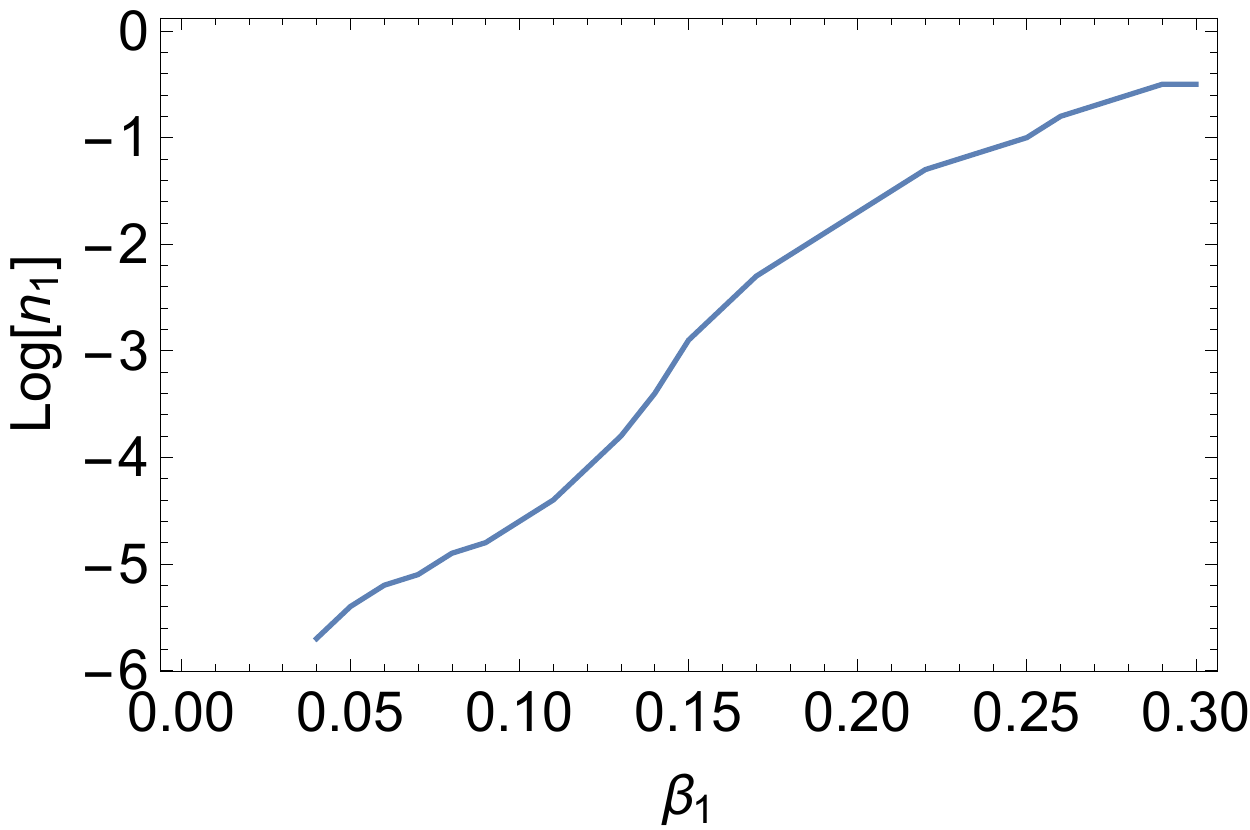}
\caption{Condition for the appearance of the isothermal jump. For a given shock velocity the isothermal  jump appears for densities above the curve.}
\label{nojump}
\end{figure}

Evolution of quantities within the shock in $\eta-\theta$ plane are pictured in Figs. \ref {eta-theta2} and \ref{eta-theta3}. Each panel is for a given velocity $\beta_1$ with different curves corresponding to different densities. 
For sufficiently small $n_1$ the isothermal jump disappears. We know that when the post-shock pressure and enthalpy are dominated by radiation, there is no isothermal jump - see \S \ref{Highly}.
Analytically this can be seen from Eq. (\ref{moment}) where the only  possibly negative term is proportional to density $n_1$.
Numerical solutions give the following conditions for the disappearance of the isothermal jump, Fig. \ref{nojump}.

\section{Limitations of the approach}

Our treatment has a number of simplifications/limitations. We treat photon propagation in diffusive approxmation, neglecting the spectral distribution and, eefectively using only the first two momenta of the photon distribution - photon density and flux. These are important limitations \citep[see, \eg][for the kinetic treatment of photons]{2018MNRAS.474.2828I}. Yet this simplification allows to elucidate  clearly the underlying  effects. We expect that far upfront the diffusive approximation will be violated
\citep[see also][ for discussion of the applicability of the diffusive approximation]{ZeldovichRaizer}.

Another important limitation is the establishment of the black-body spectrum. Using the  bremsstrahlung emissivity \citep[\eg][]{1999acfp.book.....L} the time $\tau_{phot} $ to produce enough photons can be estimated as 
\be
\tau_{phot} \approx \frac{\pi}{60} \frac{1}{\alpha_f^3}  \frac{\lambda_C}{c}  \frac{\theta^{7/2}}{ n^2} 
\ee
(here $\alpha_f $ is the fine structure constant and $n$ is a dimensionless number density normalized to $\lambda_C^3$). This time is shorter than the scattering time $\tau_{scat} = 1/( n \sigma_T c)$ for
\be
\theta \leq \alpha_f^{2/7} n^{2/7} \approx 0.3 n^{2/7}
\ee
(If number density of leptons is dominated by pairs, then $\tau_{phot}$ is always larger than $\tau_{scat}$, reaching a minimum at $\theta =1/2$, a which point 
$\tau_{phot} \approx 500 \tau_{scat}$.) The two-photon pair production can also be important \citep{2010ApJ...725...63B,2017arXiv170802633L}, yet the long times to reach the blackbody photon numbers is a limitation of the approach. 

Partly offsetting this problem is the fact that the Compton $y$-parameters on the scale of one optical depth, $y \sim 4 \theta$ is close to unity for $\theta \geq 0.1$. Also, the precursor is typically much larger than the mean free  path, see Eq. (\ref{Lp}). This implies that as the plasma is heated in the precursor,   there is a lack of photons, but  their typical energy is close to the equilibrium  with the plasma temperature. As a result the typical photon energy density, and the corresponding effect on the energy redistribution,  will be somewhat smaller.

\section{Discussion}

 In this paper we address the structure of pair- and radiation-loaded shock transitions. The first most important effect is the energy redistribution within the flow due to diffusive propagation of radiation. This generically leads to the formation of a special surface  within the flow - the isothermal jump.   (The  isothermal jump is different from highly radiative isothermal shocks; in our case the flow is energy conserving.)  The isothermal jump forms  first for plasma parameters when energy density of radiation and pair can be neglected, yet energy redistribution within the flow changes the flow properties qualitatively -  even small energy density of radiation increases the order of the governing differential equation and thus cannot be neglected.  The jump itself, on scales much smaller than the radiation scattering scale, consists of a mild shock and a narrow relation layer.
 
 For increasing post-shock temperature the role of radiation pressure  increases with respect to plasma pressure. As a result, for sufficiently strong shocks the  isothermal jump disappears - highly radiation-dominated shocks smoothly reach the final state. 
 
Overall, the  formation of pairs has only a mild effect on the shock structure in the regime $\theta \sim 1$. This is due to the fact that at sub-relativistic temperatures the number of pairs is exponentially suppressed, while at highly relativistic regime  thermodynamically pairs behave similar to photons. But formation of pairs may be important for the diffusive approximation to apply.

The parameters  of radiation-modified shocks in the wind of the NS-NS mergers,  see Eq. (\ref{param}), are  expected to be in a highly radiation-dominated regime (compare with Fig. \ref{eta-theta2}), where the isothermal jump disappears. On the other hand, the   isothermal jump could be important for core-collapse SNe: the immediate post-shock temperature could be 25\% higher than predicted by fluid jump conditions (but local density 30\% lower).  This could  influence the neutrino flux, and, perhaps,  help drive the explosion \citep[see, \eg][]{1976ApJS...32..233W}

I would like to thank Maxim Barkov and  Amir Levinson for discussions and organizers of the  workshop  ``Cosmic Accelerators'' at the Joint Space-Science Institute  where part of this work has been performed. 
This work had been supported by   NSF  grant AST-1306672, DoE grant DE-SC0016369 and NASA grant 80NSSC17K0757.

\bibliographystyle{apj}
\bibliography{/Users/maxim/Home/Research/BibTex}

\begin{thebibliography}{18}
\expandafter\ifx\csname natexlab\endcsname\relax\def\natexlab#1{#1}\fi

\bibitem[{{Abbott} {et~al.}(2017){Abbott}, {Abbott}, {Abbott}, {Acernese},
  {Ackley}, {Adams}, {Adams}, {Addesso}, {Adhikari}, {Adya}, \&
  et~al.}]{2017ApJ...848L..13A}
{Abbott}, B.~P., {et~al.} 2017, \apjl, 848, L13

\bibitem[{{Bromberg} {et~al.}(2017){Bromberg}, {Tchekhovskoy}, {Gottlieb},
  {Nakar}, \& {Piran}}]{2017arXiv171005897B}
{Bromberg}, O., {Tchekhovskoy}, A., {Gottlieb}, O., {Nakar}, E., \& {Piran}, T.
  2017, ArXiv e-prints

\bibitem[{{Budnik} {et~al.}(2010){Budnik}, {Katz}, {Sagiv}, \&
  {Waxman}}]{2010ApJ...725...63B}
{Budnik}, R., {Katz}, B., {Sagiv}, A., \& {Waxman}, E. 2010, \apj, 725, 63

\bibitem[{{Chapline} \& {Granik}(1984)}]{1984PhFl...27.1991C}
{Chapline}, G.~F., \& {Granik}, A. 1984, Physics of Fluids, 27, 1991

\bibitem[{{Gottlieb} {et~al.}(2017){Gottlieb}, {Nakar}, {Piran}, \&
  {Hotokezaka}}]{gnph17}
{Gottlieb}, O., {Nakar}, E., {Piran}, T., \& {Hotokezaka}, K. 2017, ArXiv
  e-prints

\bibitem[{{Ito} {et~al.}(2018){Ito}, {Levinson}, {Stern}, \&
  {Nagataki}}]{2018MNRAS.474.2828I}
{Ito}, H., {Levinson}, A., {Stern}, B.~E., \& {Nagataki}, S. 2018, \mnras, 474,
  2828

\bibitem[{{Landau} \& {Lifshitz}(1959)}]{LLVI}
{Landau}, L.~D., \& {Lifshitz}, E.~M. 1959, {Fluid mechanics}

\bibitem[{{Lang}(1999)}]{1999acfp.book.....L}
{Lang}, K.~R. 1999, {Astrophysical formulae}

\bibitem[{{Lazzati} {et~al.}(2017){Lazzati}, {L{\'o}pez-C{\'a}mara},
  {Cantiello}, {Morsony}, {Perna}, \& {Workman}}]{2017ApJ...848L...6L}
{Lazzati}, D., {L{\'o}pez-C{\'a}mara}, D., {Cantiello}, M., {Morsony}, B.~J.,
  {Perna}, R., \& {Workman}, J.~C. 2017, \apjl, 848, L6

\bibitem[{{Lundman} {et~al.}(2017){Lundman}, {Beloborodov}, \&
  {Vurm}}]{2017arXiv170802633L}
{Lundman}, C., {Beloborodov}, A., \& {Vurm}, I. 2017, ArXiv e-prints

\bibitem[{{Metzger} {et~al.}(2010){Metzger}, {Mart{\'{\i}}nez-Pinedo},
  {Darbha}, {Quataert}, {Arcones}, {Kasen}, {Thomas}, {Nugent}, {Panov}, \&
  {Zinner}}]{2010MNRAS.406.2650M}
{Metzger}, B.~D., {et~al.} 2010, \mnras, 406, 2650

\bibitem[{{Mezzacappa} {et~al.}(2001){Mezzacappa}, {Liebend{\"o}rfer},
  {Messer}, {Hix}, {Thielemann}, \& {Bruenn}}]{2001PhRvL..86.1935M}
{Mezzacappa}, A., {Liebend{\"o}rfer}, M., {Messer}, O.~E., {Hix}, W.~R.,
  {Thielemann}, F.-K., \& {Bruenn}, S.~W. 2001, Physical Review Letters, 86,
  1935

\bibitem[{{Pozanenko} {et~al.}(2017){Pozanenko}, {Barkov}, {Minaev}, {Volnova},
  {Mazaeva}, {Moskvitin}, {Krugov}, {Samodurov}, {Loznikov}, \&
  {Lyutikov}}]{pbm17}
{Pozanenko}, A., {et~al.} 2017, ArXiv:1710.05448

\bibitem[{{Rayleigh}(1910)}]{1910RSPSA..84..247R}
{Rayleigh}, L. 1910, Proceedings of the Royal Society of London Series A, 84,
  247

\bibitem[{{Svensson}(1984)}]{1984MNRAS.209..175S}
{Svensson}, R. 1984, \mnras, 209, 175

\bibitem[{{Wandel} \& {Yahil}(1979)}]{1979A&A....72..367W}
{Wandel}, A., \& {Yahil}, A. 1979, \aap, 72, 367

\bibitem[{{Weaver}(1976)}]{1976ApJS...32..233W}
{Weaver}, T.~A. 1976, \apjs, 32, 233

\bibitem[{{Zeldovich} \& {Raizer }(2003)}]{ZeldovichRaizer}
{Zeldovich}, Y.~B., \& {Raizer }, Y.~P. 2003, {Physics of Shock Waves} (Dover
  Publications Inc.)

\end{thebibliography}

\appendix

\section{Finite Mach number: resolving the precursor}
\label{precursor}
Equation (\ref{deta1})  has a special point upfront of the shock, at $\eta=1$, where 
 the derivative  $\partial_x \eta$ diverges, implying that the very front of the shock is located at finite distance with non-analytical behavior. Finite upstream Mach number resolves this singularity, as we demonstrate next.
 
  For finite upstream temperature
\ba &&
\rho _1 v_1=\rho  v
\nn &&
p_1+\rho _1 v_1^2=p+\rho  v^2
\nn &&
v_1 \left(\frac{1}{2} \rho _1 v_1^2+w_1\right)=F_r+v
   \left(\frac{\rho  v^2}{2}+w\right)
\nn &&   
F_r = -\frac{4 \pi ^2 c^3 m_e \theta _T^3}{45 n \lambda _C^3 \sigma _T} \partial_x \theta _T
\ea

Mass and momentum conservation give
\ba &&
\eta = \frac{\pm\sqrt{\left(\theta _{T,1}+\beta _1^2 \mu \right){}^2-4 \beta _1^2 \mu  \theta _T}+\theta
   _{T,1}+\beta _1^2 \mu }{2 \beta _1^2 \mu } \rightarrow  \frac{1}{2} \pm \frac{\sqrt{\beta _1^2 \mu -4 \theta _T}}{2 \beta _1 \sqrt{\mu }}
\nn &&
\frac{1}{2} \left(1-\sqrt{1-\frac{\theta _T}{\theta _{T,\max }}}\right)
\nn &&
\theta _T=\eta  \left(\theta _{T,1}-\beta _1^2 (\eta -1) \mu \right) \rightarrow \beta _1^2 (1-\eta ) \eta  \mu
\ea
where the second relations indicate $\theta _{T,1} \rightarrow 0$ limit.

Thus, there are two branches for $\eta(\theta)$ which connect at 
\ba &&
\eta_{crit}=\frac{1}{2} + \frac{\theta _{T,1}}{2 \beta _1^2 \mu }\rightarrow \frac{1}{2}
\nn &&
\theta _{T,max}= \frac{\left(\theta _{T,1}+\beta _1^2 \mu \right){}^2}{4 \beta _1^2 \mu } \rightarrow \frac{\beta _1^2 \mu }{4}
\ea

The equation for the compression ratio becomes
\be
\frac{\pi  \beta _1^5 (\gamma -1) \mu ^3}{15 {\alpha _f}^2 {n_1}^2 \lambda _C^5}
\partial_x \eta =
\frac{\beta _1^6 (\eta -1) \mu ^3 \left(-2 \gamma  \theta _{T,1}+\beta _1^2 \gamma  (\eta -1) \mu +\beta _1^2
   (\eta +1) \mu \right)}{\eta ^4 \left(\beta _1^2 (\eta -1) \mu -\theta _{T,1}\right){}^3 \left(\theta
   _{T,1}+\beta _1^2 (1-2 \eta ) \mu \right)}
\rightarrow
\frac{\gamma  (\eta -1)+\eta +1}{(1-2 \eta ) (1-\eta )^2 \eta ^4}
\label{deta}
\ee

 For finite $\theta _{T,1}$ the compression ratio far upstream,
 \be
 \eta \propto
 e^{\frac{30 \beta _1 n_1^2 z \lambda _C^5 \alpha _f^2}{\pi  (\gamma -1) \theta _{T,1}^3}},
\ee
is finite for all $x$ (upstream medium corresponds to $x<0$). For small upstream temperature the bulk of the transition is well described by the $\theta _{T,1} =0$ limit.

\section{Properties of the isothermal jump}
\label{properties}
Let us highlight the properties of the  isothermal jump in the simple,analytically treatable case, when the radiation and pair production effects can be neglected, while energy redistribution by photon diffusion is important, see \S \ref{first}.

 At the iso-thermal jump the sound speed is 
\be
c_s = \sqrt{\gamma T/m_p} = \frac{\sqrt{2} \sqrt{\gamma -1} \sqrt{\gamma }}{\gamma +1} v_1
\ee
At this point, on the upper branch the  parameters of the flow are
\ba && 
\eta_+ = 2/(\gamma+1)
\nn &&
v_+= \frac{2}{\gamma +1} v_1 = \frac{3}{4} v_1
\nn &&
M_+ = \frac{\sqrt{2}}{\sqrt{\gamma -1} \sqrt{\gamma }} = \frac{3}{\sqrt{5}}
\ea
While in the post-jump flow
\ba
 && 
\eta_2 = \frac{\gamma-1}{\gamma+1}
\nn &&
v_2= \frac{\gamma+1}{\gamma-1}v_1 = \frac{1}{4} v_1
\nn &&
M_2 = \frac{\sqrt{\gamma -1}}{ \sqrt{2 \gamma }} = \frac{1}{\sqrt{5}}
\ea
Thus, the compression ratio at the  isothermal jump is $ (\gamma-1)/2=1/3$.

\end{document}